\documentclass[prd,aps,onecolumn,tightenlines,notitlepage,nofootinbib,preprintnumbers,letterpaper,superscriptaddress]{revtex4-1}  

\pdfoutput=1
\usepackage{epsfig}
\usepackage{amsfonts}
\usepackage{amsmath}
\usepackage{graphicx}
\usepackage{color}
\usepackage{xcolor}
\usepackage{amssymb,amsmath,hyperref,slashed,color,graphicx,bm, url}
\usepackage[T1]{fontenc}
\usepackage{relsize}
\usepackage{marvosym}
\usepackage{booktabs,multirow} 
\usepackage{accents}
\usepackage{mathrsfs}
\usepackage{wasysym}

\DeclareMathAlphabet{\mathpzc}{OT1}{pzc}{m}{it}
\DeclareMathAlphabet{\mathbf}{U}{bf}{m}{n}
\DeclareMathAlphabet{\mathfrak}{U}{frak}{m}{n}

\newcommand{\cevns}{CEvNS}

\date{March 2021}

\begin{document}

\preprint{FERMILAB-PUB-21-357-T, NUHEP-TH/21-08, IFT-UAM/CSIC-21-92}

\title{The physics potential of a reactor neutrino experiment with Skipper-CCDs: \\
Searching for new physics with light mediators}

\author{G. Fernandez-Moroni}
\email[E-mail:]{gfmoroni@fnal.gov}
\thanks{\scriptsize \!\! \href{https://orcid.org/0000-0002-1654-9562}{0000-0002-1654-9562}}
\affiliation{Fermi National Accelerator Laboratory, Batavia, IL, 60510, USA}

\author{R.~Harnik}   
\email[E-mail:]{roni@fnal.gov}
\thanks{\scriptsize \!\! \href{https://orcid.org/0000-0001-7293-7175}{0000-0001-7293-7175}}
\affiliation{Fermi National Accelerator Laboratory, Batavia, IL, 60510, USA}

\author{P.~A.~N.~Machado}   
\email[E-mail:]{pmachado@fnal.gov}
\thanks{\scriptsize \!\! \href{https://orcid.org/0000-0002-9118-7354}{0000-0002-9118-7354}}
\affiliation{Fermi National Accelerator Laboratory, Batavia, IL, 60510, USA}

\author{I. Martinez-Soler}   
\email[E-mail:]{ivan.martinezsoler@northwestern.edu}
\thanks{\scriptsize \!\! \href{https://orcid.org/0000-0002-0308-3003}{0000-0002-0308-3003}}
\affiliation{Fermi National Accelerator Laboratory, Batavia, IL, 60510, USA}
\affiliation{Northwestern University, Department of Physics \& Astronomy, 2145 Sheridan Road, Evanston, IL 60208, USA}
\affiliation{Colegio de F\'isica Fundamental e Interdisciplinaria de las Am\'ericas (COFI), 254 Norzagaray street, San Juan, Puerto Rico 00901.}

\author{Y.~F.~Perez-Gonzalez}   
\email[E-mail:]{yfperezg@northwestern.edu}
\thanks{\scriptsize \!\! \href{https://orcid.org/0000-0002-2020-7223}{0000-0002-2020-7223}}
\affiliation{Fermi National Accelerator Laboratory, Batavia, IL, 60510, USA}
\affiliation{Northwestern University, Department of Physics \& Astronomy, 2145 Sheridan Road, Evanston, IL 60208, USA}
\affiliation{Colegio de F\'isica Fundamental e Interdisciplinaria de las Am\'ericas (COFI), 254 Norzagaray street, San Juan, Puerto Rico 00901.}

\author{D. Rodrigues}
\email[E-mail:]{rodriguesfm@df.uba.ar}
\thanks{\scriptsize \!\! \href{https://orcid.org/0000-0002-7952-7168}{0000-0002-7952-7168}}
\affiliation{Fermi National Accelerator Laboratory, Batavia, IL, 60510, USA}
\affiliation{Department of Physics, FCEN, University of Buenos Aires and IFIBA, CONICET, Buenos Aires,
Argentina}

\author{S. Rosauro-Alcaraz}   
\email[E-mail:]{salvador.rosauro@uam.es}
\thanks{\scriptsize \!\! \href{https://orcid.org/0000-0003-4805-5169}{0000-0003-4805-5169}}
\affiliation{Departamento de F\'isica T\'eorica and Instituto de F\'isica T\'eorica, IFT-UAM/CSIC, Universidad Aut\'onoma
de Madrid, Cantoblanco, 28049, Madrid, Spain}

\begin{abstract}
We explore the sensitivity to new physics of the recently proposed
vIOLETA experiment: a 10~kg Skipper Charged Coupled Device detector
deployed 12 meters away from a commercial nuclear reactor core.  We
investigate two broad classes of models which benefit from the very
low energy recoil threshold of these detectors, namely neutrino
magnetic moments and light mediators coupled to neutrinos and quarks
or electrons.  We find that this experimental setup is very sensitive
to light, weakly coupled new physics, and in particular that it could
probe potential explanations of the event excess observed in XENON1T.
We also provide a detailed study on the dependence of the sensitivity
on the experimental setup assumptions and on the neutrino flux
systematic uncertainties.

\end{abstract}

\maketitle

\section{Introduction}

The recent observation~\cite{Akimov:2017ade} of coherent
neutrino-nucleus elastic scattering~\cite{Freedman:1973yd}, or CEvNS
for short, provides a novel window to probe standard and beyond
standard physics in the neutrino sector.  A key feature of CEvNS is
that it only takes place when the momentum transferred to the nucleus
is relatively small, near or below the MeV scale, so that the neutrino
can coherently scatter off the whole nucleus instead of distinguishing
its individual nucleons.  The coherence of the interaction results in
sensitivity to the square of the total weak charge of the nucleus,
which could enhance the coherent scattering cross section by a factor
10-100 when compared to neutrino-nucleon scattering. In addition, new
non-standard interactions of neutrinos with matter can be enhanced at
low recoil energies if the new interaction is mediated by a light
particle~\cite{Harnik:2012ni}. The light mediator can be a new beyond
standard model (BSM) particle, such as a $B-L$ gauge boson or a light
scalar, but it can also be the photon, if the neutrino possesses a
magnetic dipole moment. Both possibilities would signal the presence
of new BSM physics.

While the low momentum transfer results in the so-called coherent
enhancement, which is a unique advantage of CEvNS, it also poses the
main challenge in leveraging this new interaction channel.  Detecting
low energy nuclear recoils and distinguishing it from backgrounds such
as those induced by cosmic rays or neutron-nucleus scattering is a
difficult experimental task. As a result, CEvNS has yet to be
conclusively observed~\cite{Colaresi:2021kus} with reactor neutrinos
due to their low energy, despite reactors being the most intense
artificial source of neutrinos.

A novel technology that has yet to be exploited for CEvNS is that of
Skipper charged coupled devices, or Skipper-CCDs.  A basic explanation
of the Skipper-CCD technology can be found in our previous
work~\cite{Fernandez-Moroni:2020yyl}, but we highlight here two key
features of these detectors: (1) Skipper-CCDs are endowed with
single-electron counting capability~\cite{Tiffenberg2017,
  skipper_2012}, which allows them to observe energy deposits as low
as a few electronvolts; and (2) the readout time for this technology
is relatively large, about 1 millisecond per 15 $\mu$m by 15 $\mu$m
pixel, which makes impractical to use active vetos in order to reduce
cosmic ray backgrounds.

The low thresholds of Skipper-CCDs are of major importance for two
reasons.  First, detectors with very low recoil thresholds can probe
lower energy neutrinos whose interaction is not affected by possible
loss of coherence in CEvNS due to larger momentum transfer.  The loss
of coherence with increasing momentum transfer is typically
parametrized by form factors which may contain large uncertainties in
the coherent-incoherent transition region.  Thus, focusing on low
nuclear recoils may help in controlling form factor
systematics\footnote{Nevertheless, other uncertainties are relevant in
  the low recoil energy region, particularly those related to the
  fraction of nuclear recoil energy that goes into
  ionization~\cite{Fernandez-Moroni:2020yyl}, that is, the quenching
  factor; and those related to other phenomena such as the Migdal
  effect~\cite{Ibe:2017yqa, Bell:2019egg, Knapen:2020aky,
    Liao:2021yog}.}. Second, a low energy threshold allows for using
the most intense artificial source of neutrinos known, antineutrinos
from nuclear reactors.

In this work, we will estimate the sensitivity to new physics in the
neutrino sector of an experimental design similar to the recently
proposed \emph{Neutrino Interaction Observation with a Low Energy
  Threshold Array} (vIOLETA) experiment: a 10~kg Skipper-CCD detector
placed 12 meters away from a commercial nuclear reactor core.
Although a vast body of work is available in beyond standard model
(BSM) physics searches using CEvNS or related
experiments~\cite{Coloma:2017ncl, Farzan:2017xzy, Liao:2017uzy,
  Bertuzzo:2017tuf, Esteban:2018ppq, AristizabalSierra:2018eqm,
  Altmannshofer:2018xyo, Abdullah:2018ykz, Gonzalez-Garcia:2018dep,
  Giunti:2019xpr, AristizabalSierra:2019ufd, Bischer:2019ttk,
  Canas:2019fjw, Babu:2019mfe, Denton:2020hop, Flores:2020lji,
  Dent:2019ueq, Pospelov:2011ha, Harnik:2012ni, Pospelov:2012gm,
  Kosmas:2017tsq, Farzan:2018gtr, Boehm:2018sux, Denton:2018xmq,
  Cadeddu:2018dux, Billard:2018jnl, Dutta:2019eml, Cadeddu:2020nbr,
  Billard:2014yka, Cerdeno:2016sfi, Amaral:2020tga, Formaggio:2011jt,
  Anderson:2012pn, Dutta:2015nlo, Canas:2017umu, Kosmas:2017zbh,
  Blanco:2019vyp, Miranda:2020syh, Cui:2017ytb, Ge:2017mcq,
  Bertuzzo:2018itn, Dutta:2019nbn}, there is not much work done in the
direction of characterizing the physics reach of a setup like vIOLETA.

Neutrino-electron scattering, on the other hand, is a well known
process that nowadays serves as a standard candle for different
experiments~\cite{MINERvA:2019hhc,Marshall:2019vdy}.  One important
advantage that measurements of neutrino-electron interaction have over
CEvNS is the lack of dependence on the quenching factor.
Nevertheless, the cross-section for this process is several orders of
magnitude smaller than neutrino-nuclei cross sections.  Still, it
offers a great environment to search for BSM physics, specially if
such physics hides in low energy
regimes~\cite{Deniz:2009mu,Beda:2013mta,Ballett:2019xoj,Aprile:2020tmw}.
In fact, the XENON1T collaboration presented recently results related
to searches of low recoil electrons finding an excess that could be
explained if neutrinos have a non-zero magnetic
moment~\cite{Aprile:2020tmw}.  An experiment like vIOLETA, having a
noticeably low threshold to observe electrons, could measure
neutrino-electron interactions at small momentum transfer regimes,
thus being able to test different BSM scenarios.  As illustrative
cases, we will analyze the experimental sensitivity to non-standard
neutrino magnetic moment, specially to the XENON1T hint, and light
mediators coupling to neutrinos and quarks or electrons.
Interestingly, since both neutrino-electron and CEvNS interactions
have the same experimental signature in a Skipper-CCD, there will be a
non-trivial interplay between them in the sensitivities that will be
derived for vIOLETA.

This work is organized as follows. In Sec.~\ref{sec:models}, we
establish our theoretical framework, and describe in some detail the
BSM models that we will undertake.  We give special emphasis to the
magnetic moment, in order to be as self-contained as possible.
Sec.~\ref{sec:SCCD} details the key properties of the Skipper-CCD
technology that will de the basis of our sensitivity
studies. Secs.~\ref{sec:analysis} and~\ref{sec:results} we present the
procedure and results of our analysis of new light physics.  Finally,
we present our conclusions in Sec.~\ref{section:conclusions}.  We use
natural units where $\hbar = c = k_{\rm B} = 1$ throughout this
manuscript, unless otherwise stated.

\section{Searching for Physics Beyond the Standard Model}\label{sec:models}

Given the large statistics expected in a reactor experiment using the Skipper-CCD technology --${\cal O}(10^5)$ CEvNS events for a 2~GW$_{\rm th}$ reactor, and 3 kg-year exposure--, we can anticipate improvements on existing constraints on physics beyond the Standard Model. Here, we will focus on two main categories for such BSM scenarios: first, models which alter specific neutrino properties, such as its magnetic moment; and second, additional low-energy interactions mediated by new, light 
degrees of freedom which impact the scattering rate of neutrinos with the targets on the detector.

In order to grasp the effect of BSM scenarios, let us first establish our notation by formulating the SM cross sections for the relevant scatterings that will be considered, neutrino-electron elastic scatterring and CEvNS.
\begin{itemize}

\item \emph{Electron antineutrino-electron scattering.} Neutrino-electron elastic scattering is one of the most relevant and well-known neutrino scattering channels. Since the reactor flux is composed only by electron antineutrinos, let us first record the differential cross section for $\overline{\nu}_e+e^-\to \overline{\nu}_e+e^-$ in terms of the electron recoil energy $E_R$,
 \begin{align}
	\left.\frac{d\sigma_{\nu e}}{dE_R}\right|_{\rm SM}&=\frac{2G_F^2m_e}{\pi}\left[\sin^4\theta_W+\left(\frac{1}{2}+\sin^2\theta_W\right)^2\left(1-\frac{E_R}{E_\nu}\right)^2-\sin^2\theta_W\left(\frac{1}{2}+\sin^2\theta_W\right)\frac{m_e E_R}{E_\nu^2}\right],\notag\\
	&\approx 4.3\times 10^{-51}\left[4\sin^4\theta_W+\left(1+2\sin^2\theta_W\right)^2\left(1-\frac{E_R}{E_\nu}\right)^2\right]\frac{\rm cm^2}{\rm eV}
\end{align}
where $G_F$ is the Fermi constant, $\theta_W$ is the weak mixing angle, $m_e$ the electron mass, and $E_\nu$ the incoming neutrino energy. In the following, we consider the sine squared of the weak mixing angle to be $\sin^2\theta_W=0.238$, value obtained from the $\overline{\rm MS}$ renormalization scheme~\cite{Zyla:2020zbs}.

\item \emph{\cevns.} Coherent scattering is a purely neutral current process where a neutrino (or antineutrino in our case) elastically scatters off  a nucleus, producing a small recoil which is the experimental signature for these events. 
The differential cross section as function of the recoil energy is
\begin{align}
	\left.\frac{d\sigma_{\rm CEvNS}}{dE_R}\right|_{\rm SM}&=\frac{G_F^2m_N}{4\pi}\left({Q}_V^{\rm SM}\right)^2\left(1-\frac{m_N E_R}{2E_\nu^2}-\frac{E_R}{E_\nu}\right)\mathcal{F}^2(E_R)\notag\\
	&\approx 2.3\times 10^{-38}\left(\frac{{Q}_V^{\rm SM}}{14}\right)^2\left(\frac{m_N}{28m_p}\right)\left(1-\frac{m_N E_R}{2E_\nu^2}\right)\frac{\rm cm^2}{\rm eV}\label{eq:cevns}
\end{align}
where ${Q}_V^{\rm SM}=N+(4\sin^2\theta_W-1)Z$ corresponds to the SM weak coupling between the nucleus and neutrinos, $N$, $Z$ are the number of neutrons and protons; and $\mathcal{F}^2(E_R)$ is a nuclear form factor. 
For simplicity we use the Helm form factor~\cite{Engel:1991wq, Lewin:1995rx, Harnik:2012ni}
\begin{align}\label{eq:FormFactor}
{\cal F}(E_R)&=3e^{-k^2s^2/2}\left[\sin(kr)-kr\cos(kr)\right]/(kr)^3
\end{align} 
where $k\equiv\sqrt{2 m_N E_R}$, $s \simeq 1$~fm is the nuclear skin thickness, $r=\sqrt{R^2-5s^2}$, and $R\simeq 1.14 \, (Z+N)^{1/3}$ is the effective nuclear radius. 
The approximation done in the second line of Eq.~\eqref{eq:cevns}  assumes silicon as target ($N=Z=14$), $E_R\ll E_\nu$, and  $\mathcal{F}^2(E_R)\approx 1$ which holds for the energy regime of  reactor antineutrinos.

\end{itemize}

In general, the presence of any BSM physics will modify the previous cross sections, thus altering the expected number of events in a detector. In a general fashion, we write the total cross section in the presence of BSM as
\begin{align}
	\frac{d\sigma}{dE_R}= \left.\frac{d\sigma}{dE_R}\right|_{\rm SM} + \left.\frac{d\sigma}{dE_R}\right|_{\rm BSM},
\end{align}
where the first term is the SM cross section for either neutrino-electron and \cevns~interactions, and the second is the modification created by the BSM interactions. 
Note that any possible interference effect that can appear according to the nature of the new mediators are included in the BSM cross section. 

\subsection{Neutrino electromagnetic properties: general framework}

One of the simplest scenarios of BSM interactions corresponds to studying the electromagnetic properties of neutrinos. In order to consider such interactions, let us first establish our notation, and discuss the  current experimental constraints on the neutrino magnetic moment coming from solar and reactor neutrinos.  
We begin with the following effective Hamiltonian for neutrino interactions with photons~\cite{Giunti:2014ixa,Giunti:2015gga}
\begin{align}
    \mathscr{H}_{\rm eff} = j_\mu A^\mu = \overline{\nu}_i \Lambda_\mu^{ik} \nu_k  A ^\mu,
\end{align}
where $i,k$ are mass eigenstate indexes, and $\Lambda_\mu^{ik}$ is the vertex function containing the information on the electromagnetic neutrino properties. Choosing the initial and final neutrino momenta to be $p_i$ and $p_f$, with $q\equiv p_i-p_f$, allows us to write the most general parametrization of $\Lambda_\mu^{ik}(p_f,p_i)\equiv \langle \nu_i (p_f, h_f) | j_\mu  | \nu_k (p_i, h_i) \rangle$~\cite{Kayser:1982br}, as 
\begin{align}
    \langle \nu_i (p_f, h_f) | j_\mu  | \nu_k (p_i, h_i) \rangle = \overline{u}(p_f) \left\{[f_Q^{ik}(q^2) + f_A^{ik}(q^2) q^2 \gamma_5](\gamma_\mu - q_\mu \slashed{q}/q^2)-i\sigma_{\mu\nu} q^\nu[f_M^{ik}(q^2) + i f_E^{ik}(q^2)\gamma_5]\right\} u (p_i),
\end{align}
where $f_X^{ik}$, $X = \{Q,M,E,A\}$, correspond to the form factors related to electric charge, magnetic, electric and anapole moments, respectively and $\sigma_{\mu\nu}$ is defined as usual, $\sigma^{\mu\nu}=\frac{1}{2}[\gamma^\mu,\gamma^\nu]$. Notice that these form factors are matrices in the mass eigenstate space, so we can have ``diagonal'' and ``transition'' form factors.  

At zero-momentum transfer ($q^2 = 0$), the diagonal form-factors indicate the couplings between real-photons and neutrinos. Thus, we can interpret them as the neutrino electric charge, magnetic, electric, and anapole moments, respectively
\begin{align*}
   \mathtt{q}_{ik} &= f_Q^{ik}(0), \quad \mu_{ik} = f_M^{ik}(0),\\
   \varepsilon_{ik} &= f_E^{ik}(0),\quad a_{ik} = f_A^{ik}(0).
\end{align*}
Furthermore, depending on the neutrino fermionic nature, the diagonal charge, magnetic moment and electric moment form factors can vanish (Majorana) or be non-zero (Dirac). For both Dirac and Majorana cases, the transition form factors can be non-vanishing.

We have assumed so far a general parametrization of the neutrino electromagnetic interactions. Nevertheless, one may wonder if these different terms can arise from some UV complete theory. From the point of view of the SM Effective Field Theory (SMEFT), the magnetic and electric moments can be generated from high dimensional operators. As demonstrated in previous works~\cite{Bell:2005kz,Bell:2006wi,Butterworth:2019iff,Chala:2020vqp}, the specific operators will depend on the neutrino nature. In the Dirac case, after introducing additional right-handed singlet states $\nu_R^i$, the following dimension six operators generate magnetic and electric  moments~\cite{Bell:2005kz,Butterworth:2019iff,Chala:2020vqp}
\begin{align}\label{eq:DEFT}
    {\cal O}^{ij}_{RB} = \overline{L_i}\sigma^{\mu\nu}\nu_R^j \widetilde{H}B_{\mu\nu},\quad {\cal O}^{ij}_{RW} = \overline{L_i}\sigma^{\mu\nu}\nu_R^j \sigma_I \widetilde{H}W^I_{\mu\nu}\,, 
\end{align}
where $L_i, H$ are the lepton and Higgs doublets with $\widetilde{H} = i\sigma_2 H^*$, respectively, whilst $B_{\mu\nu}, W_{\mu\nu}^I$ correspond the field strengths associated to gauge bosons of the SU(2)$_L$ and U(1)$_Y$ groups, respectively, and $\sigma_I$ are the Pauli matrices. For Majorana neutrinos, and considering only SM fields, one needs to go to higher dimensions, specifically to dimension seven operators~\cite{Bell:2006wi}
\begin{align}\label{eq:MEFT}
    {\cal O}^{ij}_{B} = (\overline{L^c_i}\sigma_2 H)\sigma^{\mu\nu}(H^T \sigma_2 L^c_j)B_{\mu\nu},\quad {\cal O}^{ij}_{W} = (\overline{L^c_i}\sigma_2 H)\sigma^{\mu\nu}(H^T \sigma_2\sigma_I L^c_j)W^I_{\mu\nu}\,, 
\end{align}
$L^c_i$ being the charge conjugated lepton doublet. In the Majorana case the dipole operator is anti-symmetric, and hence a dipole interaction always leads to a flavor transition.
Whatever the neutrino nature is, a linear combination of the previous operators $C_{(R)B}^{ij}{\cal O}^{ij}_{(R)B} + C_{(R)W}^{ij}{\cal O}^{ij}_{(R)W}$, after electroweak symmetry breaking, generates magnetic $f_M^{ij}$ and electric $f_E^{ij}$ moment terms, including possible flavor transition operators. 
A neutrino millicharge, on the other hand, can arise due to the presence of additional interactions. For instance, considering anomaly-free $U(1)_{L_\alpha-L_\beta}$ scenarios, where $L$ is the lepton number and $\alpha,\beta=\{e,\mu,\tau\}$, it is possible to modify the definition of lepton hypercharges such that neutrinos obtain a non-zero charge, $Q_{\nu_\alpha}=\epsilon {\cal Q}_{\alpha}$, $\epsilon$ being a small but free parameter and ${\cal Q}_{\alpha}$ the neutrino gauge charge under the new symmetry~\cite{Babu:1992sw}. In what follows, we will focus on the phenomenology of a non-standard neutrino magnetic moment, while all other form factors will be set to zero.

\subsection{Neutrino magnetic moment}

When searching for electromagnetic properties of neutrinos experimentally, neutrino oscillations should be taken into account properly. 
For the case of the magnetic moment, neglecting all other form factors, one can define an effective parameter dependent on the initial neutrino energy $E_\nu$ and travel distance $L$, that is, a magnetic moment of the flavor eigenstate~\cite{Beacom:1999wx}
\begin{align}\label{eq:SBEmu}
    \mu_{\nu_\alpha}^2 (L,E_\nu) &= \sum_i \sum_{k,l}  U_{\alpha k} U_{\alpha l}^*\, \mu_{ik} \mu_{il}^*\, e^{-i \Delta m_{kl}^2L/2E}  \notag\\ 
    &\xrightarrow[]{L\to0}\sum_i \left| \sum_k U_{\alpha k}^* \mu_{ik} \right|^2,
\end{align}
where the second line is valid for short-baseline experiments. We see then that a reactor neutrino experiment will be sensitive to this effective magnetic moment, dependent on the magnetic moments of the mass eigenstates, weighted by the mixing matrix elements. 

For experiments with solar neutrinos, like Borexino~\cite{Borexino:2017fbd}, the effective magnetic moment, $\mu_{\astrosun}^2$, is different from the one constrained in reactor experiments, as it needs to take into account solar neutrino oscillations. 
Assuming the solar neutrino flux as an incoherent mixture, $\mu_{\astrosun}^2$ can be written in terms of the oscillation probabilities as~\cite{Borexino:2017fbd}
\begin{align}\label{eq:solmu}
    \mu_{\astrosun}^2 (E_\nu) &\approx P_{ee} \mu_{\nu_e}^2 + (1-P_{ee})(\cos^2\theta_{23}\mu_{\nu_\mu}^2+\sin^2\theta_{23}\mu_{\nu_\tau}^2),
\end{align}
where $P_{ee}=\sin^4\theta_{13}+\cos^4\theta_{13}P_{ee}^{2\nu}$, is the electron neutrino survival probability, and  $P_{ee}^{2\nu}$ is the two-flavor probability, which depends on the neutrino energy~\cite{Borexino:2017fbd}.  
For low energy solar neutrinos $P_{ee}^{2\nu}\to1-\sin^2(2\theta_{12})/2$, while for high energy $P_{ee}^{2\nu}\to\sin^2\theta_{12}$.
Current constraints on the magnetic moment are (at 90\% C.L.):
\begin{align*}
    \mu_{\astrosun} &\lesssim 2.8\times 10^{-11}\mu_B\ &\text{Borexino~\cite{Borexino:2017fbd}},\\
    \mu_{\nu_e} &\lesssim 2.9\times 10^{-11}\mu_B\ &\text{GEMMA~\cite{Beda:2013mta}},\\
    \mu_{\nu_\mu} &\lesssim 6.8\times 10^{-10}\mu_B\ &\text{LSND~\cite{Auerbach:2001wg}}.
\end{align*}
Recently, the Xenon-1T collaboration has observed an excess of low recoil energy electrons, which would be consistent with a non-zero magnetic moment, $ \mu_{\astrosun} =[1.4,2.9]\times 10^{-11}\mu_B$ at the 90\% C.L.~\cite{Aprile:2020tmw}. Using the parametrization presented in Eq.~\eqref{eq:solmu}, we can translate the XENON1T hint to a value of the magnetic moment of electron (anti)neutrinos --- fixing the oscillation parameters to their best fits --- $\mu_{\nu_e}\in [1.8,3.8]\times 10^{-11}\mu_B$ at the 90\% C.L., assuming $\mu_{\nu_{\mu}}=\mu_{\nu_{\tau}}=0$. We will show that a reactor neutrino experiment using the Skipper-CCD technology could be able to test the XENON1T hint.

The presence of a non-zero magnetic moment induces modifications to the neutrino-electron and \cevns\ interactions which can be written as (instead of ``BSM'' we use the label ``$\mu$'' to identify the neutrino magnetic moment scenario)
\begin{align}
	\left.\frac{d\sigma_{\nu e}}{dE_R}\right|_{\rm \mu}&=\alpha_{\rm EM}\mu_{\nu_e}^2\left[\frac{1}{E_R}-\frac{1}{E_\nu}\right]\\
	\left.\frac{d\sigma_{\rm CE\nu NS}}{dE_R}\right|_{\rm \mu}&=\alpha_{\rm EM}\mu_{\nu_e}^2 Z^2\left[\frac{1}{E_R}-\frac{1}{E_\nu}\right]\mathcal{F}^2(E_R)
\end{align}
where $\alpha_{\rm EM} = 1/137$ is the fine structure constant. 
Notice that for \cevns~we include the same form factor as in the SM cross section. 

\subsection{New light mediators}

The second class of BSM scenarios we consider is that of light mediators. 
In these scenarios an additional mediator is included, having couplings to neutrinos, charged leptons and quarks. In the spirit of simplified models, we assume a Lagrangian at low energies which includes terms for the new interactions with the SM fermions without specifying the gauge invariant models at high energies, 
\begin{align}
	\mathscr{L}=\,&\mathscr{L}_{\rm SM} + (g_{\nu\phi}\phi\overline{\nu}_R\nu_L+{\rm h.c.}) + g_{\nu Z^\prime}\overline{\nu}_L\gamma^\mu\nu_LZ_\mu^\prime\notag\\
	&+g_{ls}\phi\overline{\ell}\ell+g_{qs}\phi\overline{q}q -ig_{lp}\phi\overline{\ell}\gamma^5\ell-ig_{qp}\phi\overline{q}\gamma^5q\notag\\
	&+g_{lv}\overline{\ell}\gamma^\mu \ell Z_\mu^\prime+g_{qs}\overline{q}\gamma^\mu qZ_\mu^\prime +g_{la}\overline{\ell}\gamma^\mu\gamma^5 \ell Z_\mu^\prime+g_{qa}\overline{q}\gamma^\mu\gamma^5 qZ_\mu^\prime.
\end{align}
We denote a new scalar or vector boson by $\phi$ or $Z'$, respectively, and the $g$'s are all dimensionless couplings.
For each scenario, the modification of both neutrino-electron and \cevns\ cross sections will have a specific shape, possibly including interference effects. 

In the specific case of \cevns, there is an additional step; we need to translate the interactions from the quark to the nucleon level. The coherence factors related to the specific mediator are given by (see e.g. Refs.~\cite{Alarcon:2011zs, Alarcon:2012nr, DelNobile:2013sia, Hill:2014yxa, Cerdeno:2016sfi})
\begin{subequations}
	\begin{align}
		Q_V^\prime &= 3 (N+Z) g_{\nu Z^\prime} g_{qv},\\
		Q_A^\prime &= 0.3 S_N g_{\nu Z^\prime} g_{qa},\\
		Q_A &= 1.3 S_N,\\
		Q_S &= 14(N+Z) + 1.1Z,
	\end{align}
\end{subequations}
corresponding to the vector, axial, SM axial, and scalar currents, being $S_N$ the nuclear spin, $g_{\nu Z^\prime}$, $g_{qv}$, the neutrino-$Z^\prime$ and quark vector couplings, respectively. 
To avoid confusion, we define specific light mediator scenarios and we analyze them separately.
In Table~\ref{tab:Int} we summarize and compile the distinct BSM contributions to the neutrino-electron and \cevns~cross sections for each light mediator scenario, together with the non-zero couplings relevant in each case.

\begin{table}[t]
\caption{Contributions to the neutrino-electron and \cevns\ cross-sections for the different scenarios considered here. The $g_V, g_A$ are given by $g_V=\frac{1}{2}+2\sin^2\theta_W, g_A=\frac{1}{2}$~\cite{Cerdeno:2016sfi}. \label{tab:Int} }
    \centering
    \begin{tabular}{cccc}
        \toprule\toprule
       Interaction  & Non-zero couplings & $\left.\frac{d\sigma_{\nu e}}{dE_R}\right|_{\rm BSM}$ & $\left.\frac{d\sigma_{\rm CE\nu NS}}{dE_R}\right|_{\rm BSM}$ \\ \midrule\midrule
        Magnetic Moment & $\mu_{\nu_e}$ & $\alpha_{\rm EM}\mu_{\nu_e}^2\frac{E_\nu-E_R}{E_\nu E_R}$  & $\alpha_{\rm EM}\mu_{\nu_e}^2 Z^2\frac{E_\nu-E_R}{E_\nu E_R}\mathcal{F}^2(E_R)$  \\\midrule
       Scalar & $g_{\nu,\phi},g_{es},g_{qs}$ & $\frac{g_{\nu,\phi}^2 g_{es}^2E_Rm_e^2}{4\pi E_\nu^2(2E_Rm_e+m_\phi^2)^2}$  & $\frac{Q_S^2m_N^2E_R g_{\nu\phi}^2g_{qs}^2}{4\pi E_\nu^2(2E_Rm_N+m_\phi^2)^2}$  \\\midrule
       Pseudoscalar&  $g_{\nu,\phi},g_{ep},g_{qp}$  & $\frac{g_{\nu,\phi}^2 g_{ep}^2E_R^2m_e}{8\pi E_\nu^2(2E_Rm_e+m_\phi^2)^2}$  & $0$  \\ \midrule
       \multirow{2}*{Vector} &\multirow{2}*{ $g_{\nu Z^\prime},g_{ev},g_{qv}$} & $\frac{\sqrt{2} G_F m_e g_V g_{\nu Z^\prime} g_{ev}}{\pi (2E_Rm_e+m_{Z^\prime}^2)}$  & $-\frac{G_F m_N Q_V^{\rm SM} Q_V^\prime (2E_\nu^2-E_Rm_N)}{2\sqrt{2}\pi E_\nu^2(2E_Rm_N+m_{Z^\prime}^2)}$  \\  
 &  & $+\frac{g_{\nu Z^\prime}^2 g_{ev}^2m_e}{2\pi (2E_Rm_e+m_{Z^\prime}^2)^2}$ & $+\frac{Q_V^{\prime 2} m_N (2E_\nu^2-E_Rm_N)}{4\pi E_\nu^2(2E_Rm_N+m_{Z^\prime}^2)^2}$\\ \midrule
        \multirow{3}*{} &  & $-\frac{\sqrt{2} G_F m_e g_A g_{\nu Z^\prime} g_{ea}}{\pi (2E_Rm_e+m_{Z^\prime}^2)}$ & $\frac{G_F m_N Q_A Q_A^\prime (2E_\nu^2+E_Rm_N)}{2\sqrt{2}\pi E_\nu^2(2E_Rm_N+m_{Z^\prime}^2)}$ \\ 
		Axial & $g_{\nu Z^\prime},g_{ea},g_{qa}$ & $+\frac{g_{\nu Z^\prime}^2 g_{ea}^2m_e}{2\pi (2E_Rm_e+m_{Z^\prime}^2)^2}$ & $-\frac{G_F m_N Q_V^{\rm SM}Q_A^\prime E_\nu E_R}{2\sqrt{2}\pi E_\nu^2(2E_Rm_N+m_{Z^\prime}^2)}$\\
		&&& $+\frac{Q_A^{\prime 2} m_N (2E_\nu^2+E_Rm_N)}{4\pi E_\nu^2(2E_Rm_N+m_{Z^\prime}^2)^2}$\\\midrule
        \bottomrule
    \end{tabular}
\end{table}


\section{Skipper-CCD technology}\label{sec:SCCD}

Before presenting our main analysis, it is useful to describe key features of Skipper-CCDs.
These devices are 15 $\mu$m$\times$15 $\mu$m pixelated sensors developed by the Lawrence Berkeley National Laboratory (LBNL).
They are fabricated on high resistivity silicon~\cite{Holland:2003}, allowing for an active depth of each pixel of 675 $\mu$m.
Thanks to its pixel size they provide excellent spatial resolution.
The main feature of the Skipper-CCD, in comparison with normal CCD, is its capability of measuring the charge in each pixel as many times as desired in a non-destructive way~\cite{skipper_2012}. 
As a result, it is possible to reach sub-electron readout noise~\cite{Tiffenberg2017}, which essentially allows for an unambiguous charge quantification in a large range from zero up to thousand electrons~\cite{Rodrigues2020}. 
The only limitation in energy resolution is given by the inevitable silicon absorption process quantified by the Fano factor, which indicates the variance to mean ratio of the charge distribution. The Fano factor for silicon is  around $12\%$~\cite{Rodrigues2020}. 

Another important aspect of Skipper-CCDs are their relatively slow readout.
The time spent in reading the charge in each pixel can be of the order of 10 millisecond to reach 0.2 e$^-$ of readout noise, which guarantee single-carrier counting capability, i.e. the capability of counting single electrons.
This slow readout makes it impossible to use active shielding to mitigate cosmic ray backgrounds.
Thus, Skipper-CCDs need to rely on  passive shielding via overburden to significantly lower their backgrounds.
Nevertheless, since the experimental setup consists on a set of Skipper-CCD detectors near a commercial nuclear reactor, the overburden is only set by the concrete over the detector and the power plant concrete dome.
We thus adopt a conservative estimate of the background rate of 1~kdru (1 \textit{differential rate unit}, or dru, is 1 event per day-keV-kg)~\cite{ThetaWeak2020}.

One of the most important consequences of the aforementioned sub-electron readout noise achievable by these sensors is the very low detection energy threshold (see section~\ref{sec:dependence-assumptions} for a discussion on this topic).
Thus, all models whose cross-section increases as the recoil energy decreases (see Table~\ref{tab:Int}) will be benefited by the low energy detection threshold of Skipper-CCDs.

The Skipper-CCD technology has already proved its potential for light dark matter searches~\cite{Barak:2020fql}.
Its irruption in this field has been so successful that nowadays kg-scale experiment are being under construction~\cite{DAMICM} or are currently in R\&D stage~\cite{OSCURA_2020}. 
Given the similarities in requirements between neutrino and dark matter experiments it is natural to also consider to exploit Skipper-CCD capabilities for neutrino physics. 
In this context arises the Neutrino Interaction Observation with a Low Threshold Array (vIOLETA) collaboration, an ongoing effort to deploy a kg-experiment in a nuclear power plant based on this technology~\cite{neutrino-2020-poster-1, violeta}.


\section{Analysis}\label{sec:analysis}

To probe novel physics in the neutrino sector at low scales, we consider the following setup: a Skipper-CCD detector monitoring the
electron antineutrino flux produced in a commercial nuclear reactor. 
A precise determination of the
neutrino interaction rate due to the large statistics of our experimental setup, combined with a data-driven determination of the background via reactor-off data taking will allow us to be sensitive to light new physics and probe unexplored parameter space.
Our study will be based on the experimental
capabilities of the Skipper-CCD technology, which is able to resolve
ionization energies up to one or two electrons (see discussion in Section~\ref{sec:dependence-assumptions}). 

To be more precise, we consider an array of Skipper-CCD detectors, deployed at 12 meters from the center of the main core of a nuclear reactor. 
The detector is considered to have a fiducial mass of 10~kg, and the core is assumed to have a thermal power of 2~GW in steady-state operation, emitting about 10$^{28}$ electron antineutrinos per year. This scenario corresponds to the Atucha II Nuclear Power plant, located at Lima, a town $\sim$100 km away from Buenos Aires capital city in Argentina. We will consider a data taking period of 3 years \cite{Fernandez-Moroni:2020yyl}.
We assume 45 days of reactor off data taking per year, as in Ref.~\cite{Fernandez-Moroni:2020yyl}, which is crucial to determine the background.
The background rate is taken to be 1~kdru, flat in ionization energy.
The quenching factor is parametrized as in Ref.~\cite{Aguilar-Arevalo:2019zme} which is based on the measurements presented in Chavarria et al.~\cite{Chavarria:2016xsi}.
All this information is  summarized in Table~\ref{tab:benchmark}.

\begin{table}[t]
\begin{center}
\begin{tabular}{ |l | c |}\hline\hline
Detector mass  & 10~kg\\ \hline
Distance to reactor core  & 12 meters\\ \hline
Core thermal power  & 2~GW$_{\rm th}$\\ \hline
Exposure  & 3 years\\ \hline
Reactor off time  & 45 days per year\\ \hline
Background  &1~kdru, flat in $E_I$ \\ \hline
Quenching factor & Chavarria~\cite{Chavarria:2016xsi} \\ \hline\hline
\end{tabular}
\end{center}
\caption{Main characteristics of the Benchmark experimental setup considered.\label{tab:benchmark}}
\end{table}

The event spectrum expected for a given isotope $q\in\{^{235}{\rm U},^{238}{\rm U}, ^{239}{\rm Pu},^{241}{\rm Pu}\}$ in such a detector configuration, in
any interaction framework, is based on the convolution of the
reactor flux with the neutrino cross section, that is
\begin{align}\label{eq:signal}
  n_{a}^q = \frac{W T}{\sum_{q'} (f_{q'} e_{q'})}\int_\textrm{bin $a$} \!\!\!\!\!\!\!\!dE_I &\left\{ \frac{N^{\text{N}}}{Q(E_I)}\left(1-\frac{E_I}{Q(E_I)}\frac{dQ(E_I)}{dE_I}\right) \int\!\! dE_\nu\, \frac{d\phi_{\bar{\nu}_e}^q}{d E_\nu}\left.\frac{d\sigma_{N}}{dE_R}\right|_{E_R=E_I/Q(E_I)}\right.\nonumber\\
  &+\left. \sum_{s} N^{e}_{s} \int\!\! dE_\nu\, \frac{d\phi_{\bar{\nu}_e}^q}{d E_\nu}\left.\frac{d\sigma_{e}}{dE_R}\right|_{E_R=E_I + E^{\text{bind}}_{s}}\right\}.
\end{align}
We have included the neutrino interaction with the nucleus
$\sigma_{N}$ and with the electrons ($\sigma_{e}$) in each energy
shell $s$. In the standard scenario, the number of events is largely
dominated by Coherent neutrino-nucleus scattering (CEvNS). $N^{\rm N}$ and
$N^{e}_{s}$ correspond to the number of targets for each
interaction. In a commercial nuclear reactor, the anti-neutrino flux
($\phi^{q}_{\bar{\nu}_{e}}$) is produced by the fission of four
different isotopes in
the energy range spanning from $\sim1-10$~MeV. 
In this
analysis, we will assume the flux follows the theoretical estimate
from Refs.~\cite{Mention:2011rk,Huber:2011wv} for energies above the
inverse beta decay threshold and the flux estimate of Vogel and
Engel~\cite{Vogel:1989iv} for neutrino energies below $1.8$~MeV. 
We distribute the event rate in ionization energy bins of 3.75~eV, corresponding
to single electron energy resolution. The
minimum ionization energy threshold considered is 15~eV, which corresponds on
average to four electrons. 
Below that energy threshold, the signal
would be affected by additional on-chip noise sources (leakage
current, spurious charge, etc.) that could produce fake
events. 
Note that the 15~eV energy threshold chosen to reduce the background is a very conservative estimate. 
Later, we will study the impact of this limit on the sensitivity.

The relation between the ionization energy ($E_{I}$) and the nuclear recoil energy ($E_{R}$) is given by the quenching factor $Q(E_{I}) =E_{I}/E_{R}$. For silicon, we will use the parametrization from Ref.~\cite{Aguilar-Arevalo:2019zme} of the measurement performed in sub-keV nuclear recoils with a neutron source~\cite{Chavarria:2016xsi}. For the interaction with electrons, the relation between the ionization and the recoil energies is given by electron binding energy $E_{I} = E_{R} - E^{\text{bind}}_{s}$, and therefore it depends on the energy shell $s$. Nevertheless, when the energy of the photons emitted after the de-excitation of the ionized
atom is lower than $\sim$1~keV (corresponding to an attenuation length in Si of $\sim$3~$\mu$m, much smaller than the pixel size), it can be safely
assumed that the energy of the photon will be absorbed close enough to the original interaction and contribute to the same cluster of pixels. 
Therefore, the total ionization energy in the detector 
will also include the binding energy of the electrons. 
For all the photons originated in electrons apart from the inner shell we can safely assume
that the recoil and the ionization reconstructed energies coincide ($E_{I} = E_{R}$).

If the energy of the emitted photon is higher than about 1~keV, its probability to travel further in the material is larger, likely not depositing its energy on the same or adjacent pixels than the original ionization due to the electrons. 
In this case, the binding energy will not contribute to the event and the reconstructed energy becomes the transferred energy of the neutrino minus the binding energy. This is relevant particularly to the peak produced by $\sim$1.8~keV X-Rays originated in the $K$-shell of silicon. As a matter of fact, the 1.8~keV peak has been observed experimentally and it can be used for calibration purposes~\cite{Aguilar-Arevalo:2019jlr}. Moreover, the peak for the total energy release when they escape the detector (total energy minus 1.8~keV) is also observed when more energetic X-rays are used to irradiate the sensor. As a conservative approach, we will treat any electron binding energy above 1~keV as missing energy.

The main background sources that will affect this measurement are Compton scattering of high energy photons and interactions of high energy neutrons produced in the atmosphere by spallation induced by muons. Although both process are well understood, a detailed analysis including the detector configuration and the different passive shielding that can be used to reduce the background is still not available. In this analysis we will assume a baseline background of 1~kdru.
This number is based on the current $\sim10$~kdru background in the CONNIE experiment and studies performed~\cite{Fernandez-Moroni:2020abn} which show that detector processing, in particular treatment to avoid partial charge collection  in the backside of the sensor, can significantly reduce backgrounds.

As a reference, we will consider a conservative minimum ionization energy of 15~eV, corresponding to an average of four ionized electrons (see discussion in section~\ref{sec:dependence-assumptions}).
Later, we will study the impact of the ionization energy threshold on the experimental sensitivity to the BSM scenarios under consideration in this work. 

To determine the sensitivity to the neutrino cross section, we will
minimize the following $\chi^2$ function
\begin{align}\label{eq:chi2}
    \chi^2=\sum_{a}\frac{\left(\mathbf{d}_{a}-\mathbf{t}_{a}\right)^2}{\mathbf{t}_{a}}+\left(\frac{\alpha^B}{\sigma^B}\right)^2+\left(\frac{\alpha_W}{\sigma_W}\right)^2+\sum_q^\textrm{isotopes}\left(\frac{\alpha_q}{\sigma_q}\right)^2,
\end{align}
where the first term accounts for the deviation of the expected number
of events between the new neutrino interaction scenario
$\mathbf{t}_a$ and the test hypothesis, the standard model case, $\mathbf{d}_a$. 
In each
ionization bin, these vectors include the contribution of the four
isotopes weighted by its fission fraction and the expected background
in each ionization bin. For the reactor configuration that we are
considering, the fission fractions used are $f_{235} : f_{238} :
f_{239} : f_{241} = 0.55 : 0.07 : 0.32 : 0.06$~\cite{Fernandez-Moroni:2020yyl}.

The statistical significance induced by the new event distribution
will be diluted due to the systematical uncertainties. In this work,
we have considered the uncertainties related to the normalization of
the background ($\sigma^{B}$), the reactor power ($\sigma_{W}$) and
the fission fraction for each isotope ($\sigma_{q}$). The deviation of
each parameter from its nominal value is parameterized by the
$\alpha$s, following the same prescription as in
\textit{Fernandez-Moroni et al.}~\cite{Fernandez-Moroni:2020yyl}. The
uncertainties associated to each parameter are given in
Table~\ref{tab:systematics}. Note that in the case of the fission
fraction, the sum must always add up to 1 ($\sum_{q}
(1+\alpha_{q})f_{q} = 1$). The background normalization will be statistically determined during the reactor-off periods. Assuming 45 days reactor-off per year, at a background rate of 1~kdru, the precision on the background will reach $1.2\%$ after 3 years. For the reactor power and the fission rate, we are taking a conservative uncertainty of $5\%$ in both cases.

\begin{table}[t]
\begin{center}
\begin{tabular}{ |l | c | c | c | c |}\hline\hline
Systematic~   & ~Background~ & ~Reactor power~& ~Relative rate per fission~ \\ \hline
Symbol   &   $\sigma_{a}^B$   &   $\sigma_{W}$   &   $\sigma_{q}$   \\
Value   &   $1.2\%$   &   $5\%$   &   $5\%$  \\ \hline \hline
\end{tabular}
\end{center}
\caption{Systematic uncertainties. See sections IV for details.\label{tab:systematics}}
\end{table}

\section{Results}\label{sec:results}

We now present the results of our analysis for the different BSM scenarios described in Section~\ref{sec:models}, according to the experimental setup summarized in Table~\ref{tab:benchmark}. 
In the following, we will be comparing the resulting sensitivity of vIOLETA to other relevant experimental results.

\subsection{Expected signal}

\begin{figure}[h!]
  \begin{center}
    \begin{tabular}{cc}
      \includegraphics[width=7.5cm]{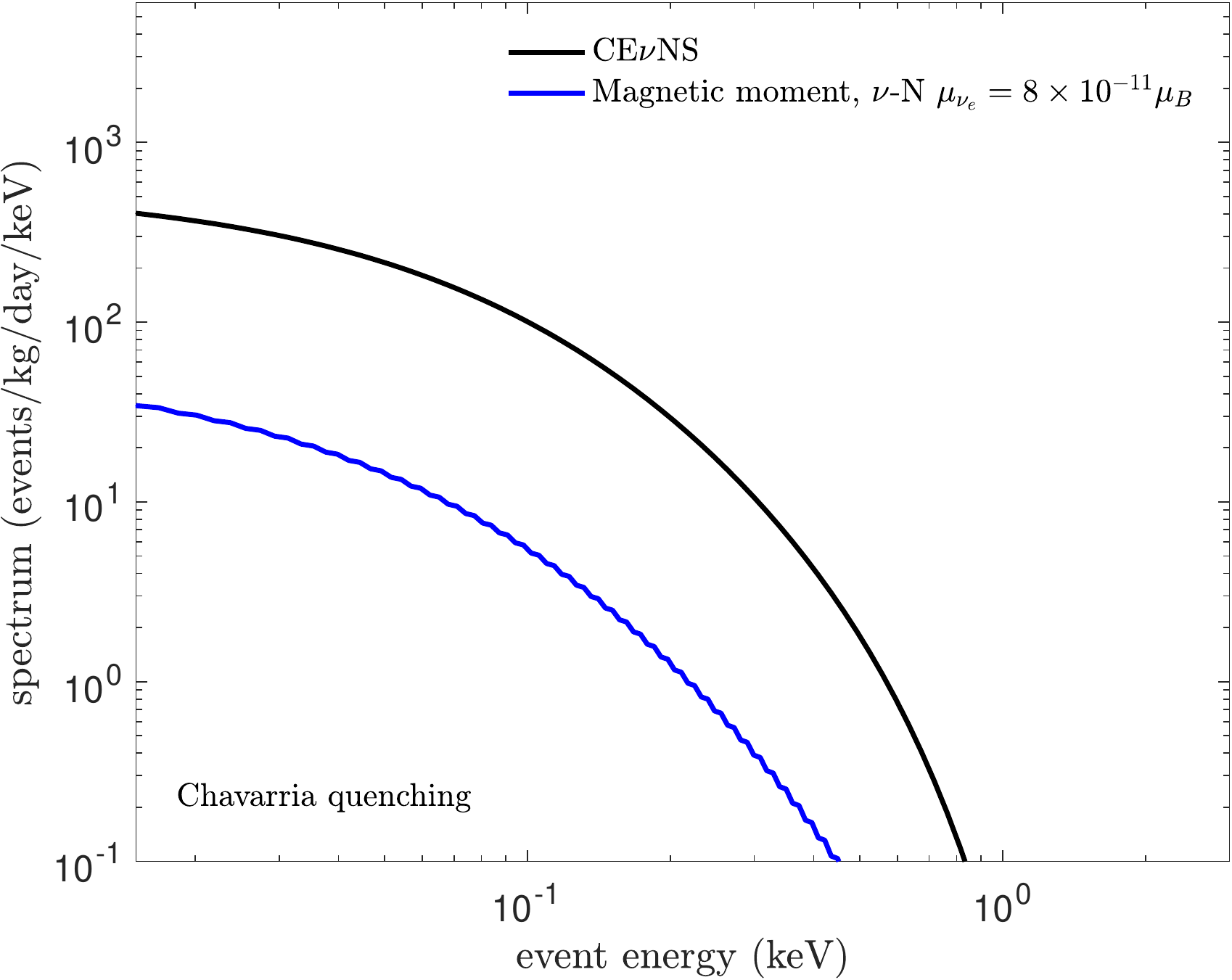} & \includegraphics[width=7.5cm]{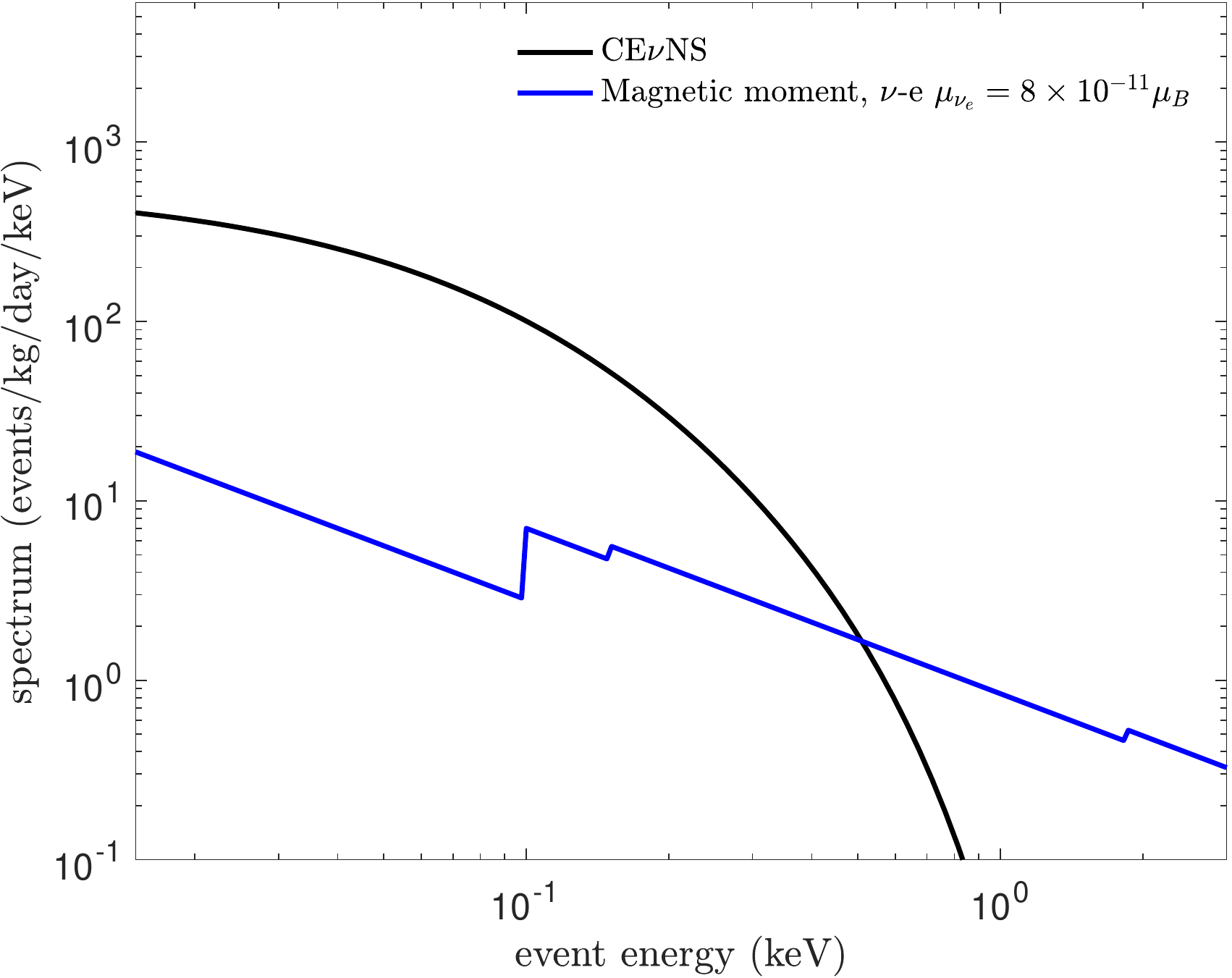}
       \\
      (a) & (b) \\[0.3cm]
      \includegraphics[width=7.5cm]{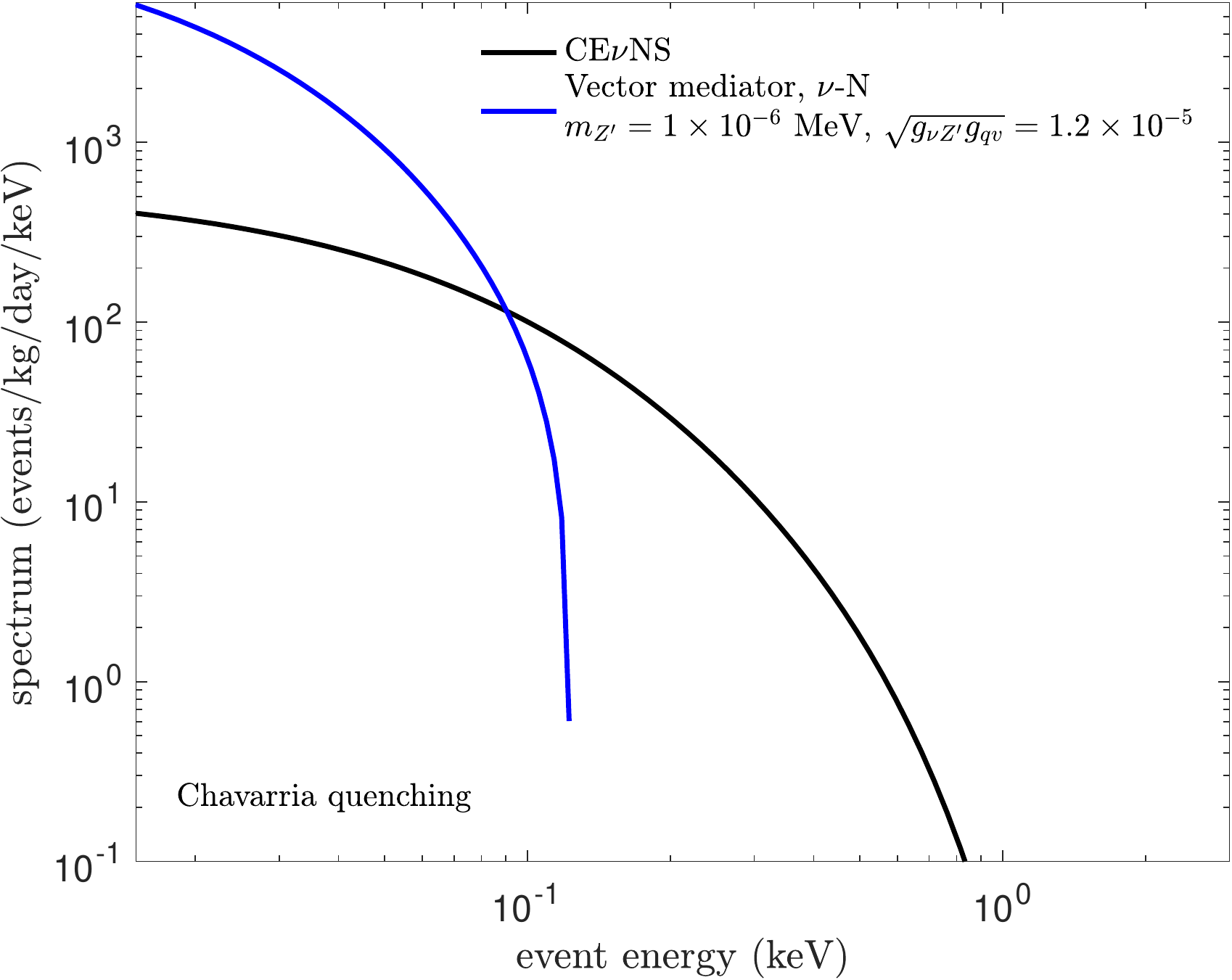} & 
      \includegraphics[width=7.5cm]{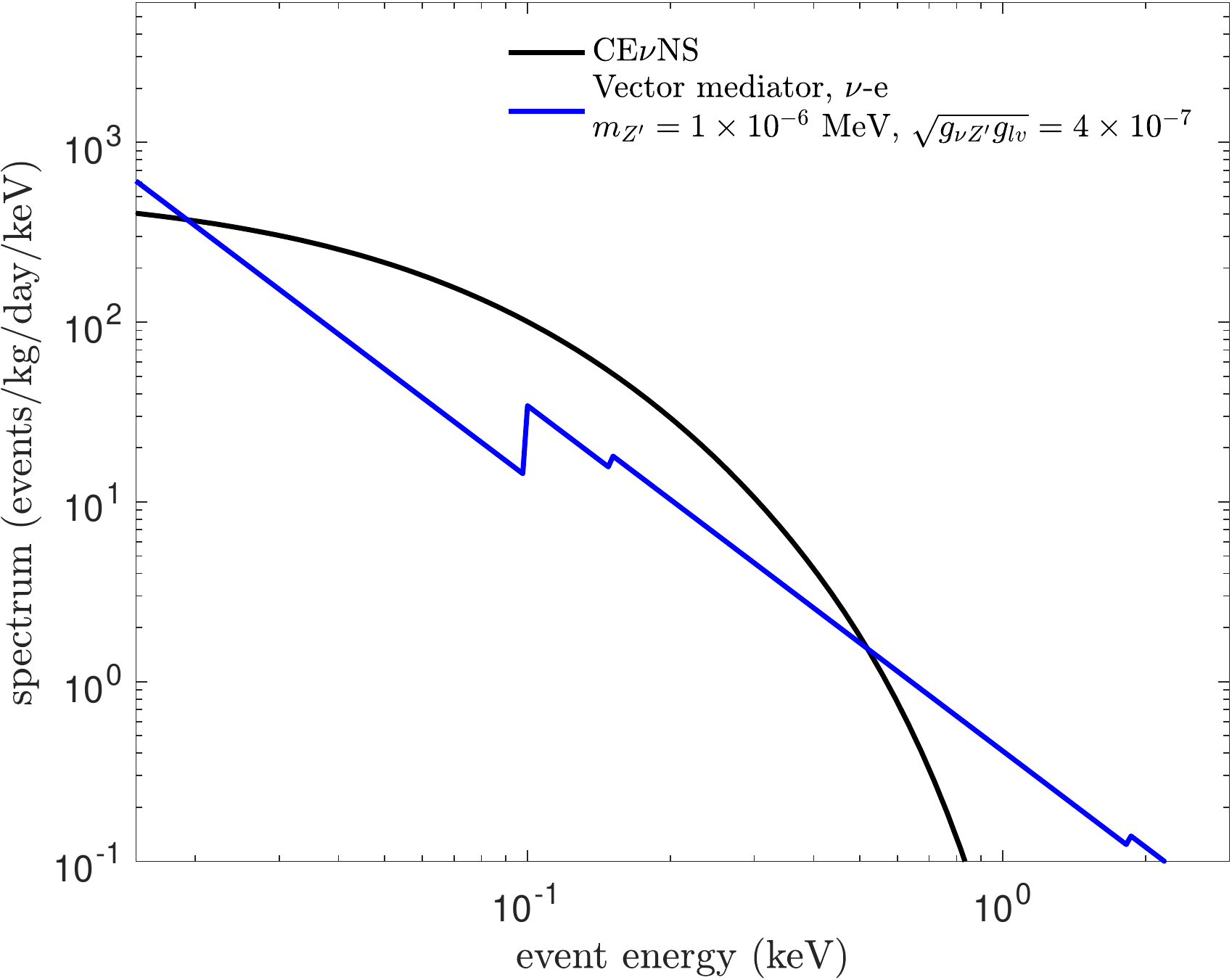}\\
      (c) & (d) \\[0.3cm]
      \includegraphics[width=7.5cm]{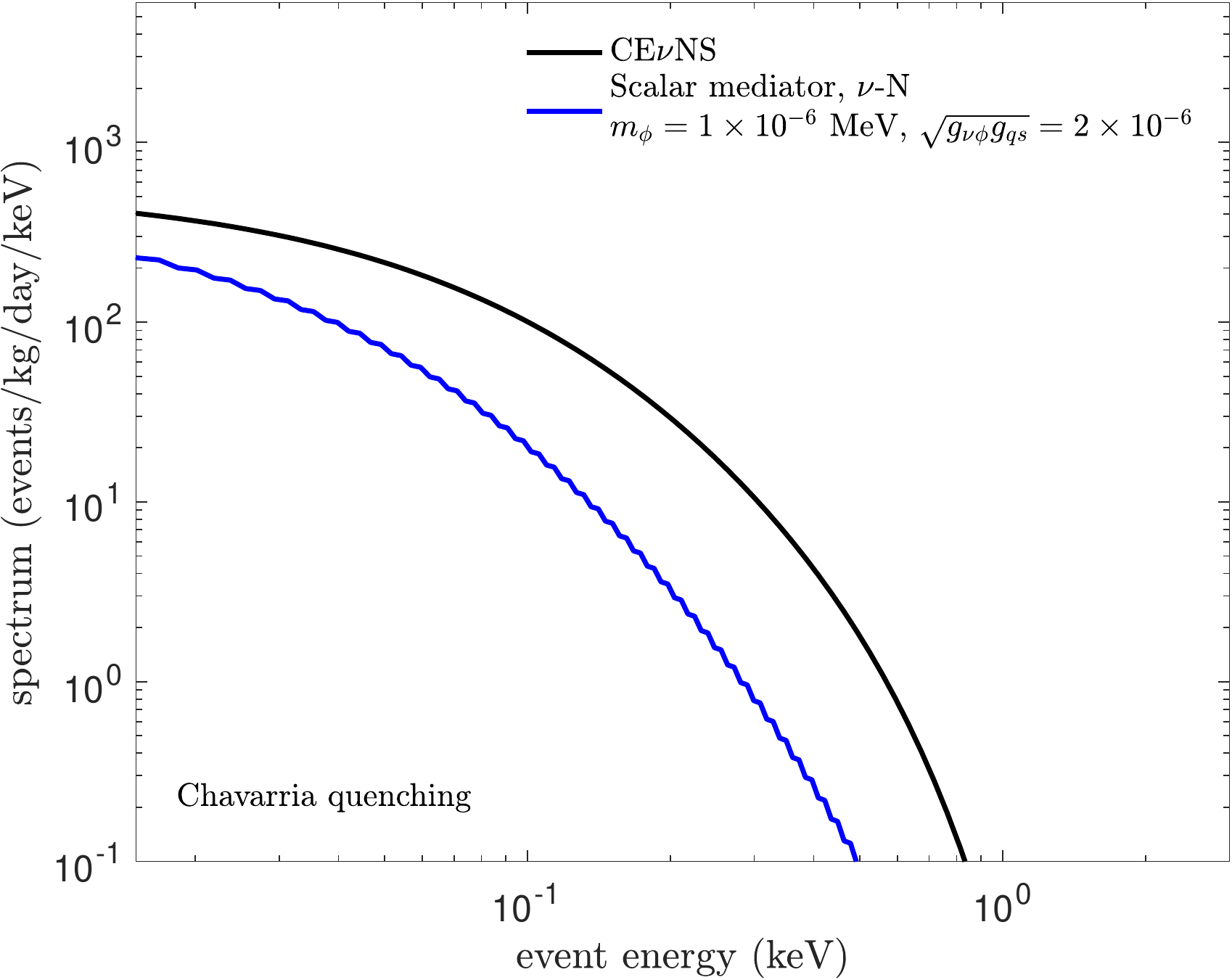} & 
      \includegraphics[width=7.5cm]{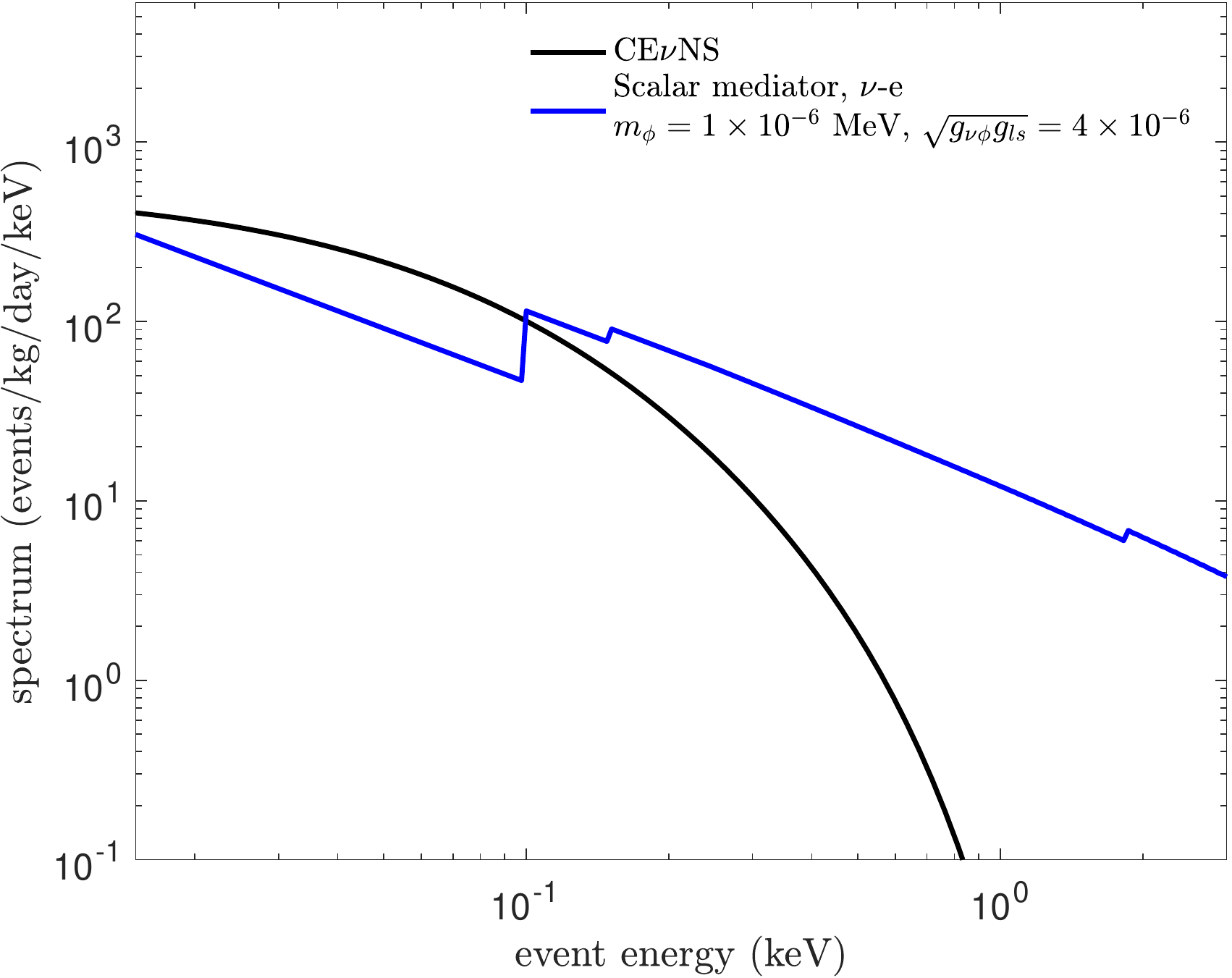} \\
      (e) & (f)
    \end{tabular}
  \end{center}
  \caption{Expected event spectra in silicon for different interaction and scattering channels, as indicated in the panels. Representative values of the parameter space region excluded by the proposed experiment are used. Spectra from nuclear interaction are shown after quenching factor. }
    \label{fig:Zprime-nuc}
\end{figure}

Before presenting the forecasted experimental sensitivities, it is useful to understand what vIOLETA would actually expect to measure if any of these BSM scenarios are realized in nature. Figure~\ref{fig:Zprime-nuc}  exhibits ionization energy distributions for some benchmarks of all models presented in section~\ref{sec:models}: neutrino magnetic moment (top panels); light vector  mediator (middle panels); and light scalar mediators (lower panels). Panels on the left exhibit the energy spectrum for nuclear recoils, while panels on the right are for electron recoils. For all benchmarks chosen, we can see that the new physics signal is significantly enhanced at low recoils, which shows the relevance of the low energy threshold of Skipper-CCD detectors when probing these BSM scenarios. For the conservative quenching factor parametrization of \emph{Chavarria et al.}~ \cite{Chavarria:2016xsi}  that we adopt here, there is an effective cutoff at recoil energies below about 400~eV. While there is no quenching factor for electrons, binding energies above 1~keV will be missing energy, for the reasons discussed previously. Therefore, the observed energy may yet be reduced with respect to the electron recoil energy. This reduction is responsible for the kinks in the left panels of Figure~\ref{fig:Zprime-nuc}.

\begin{figure}[h!]
    \includegraphics[width = 0.45\textwidth]{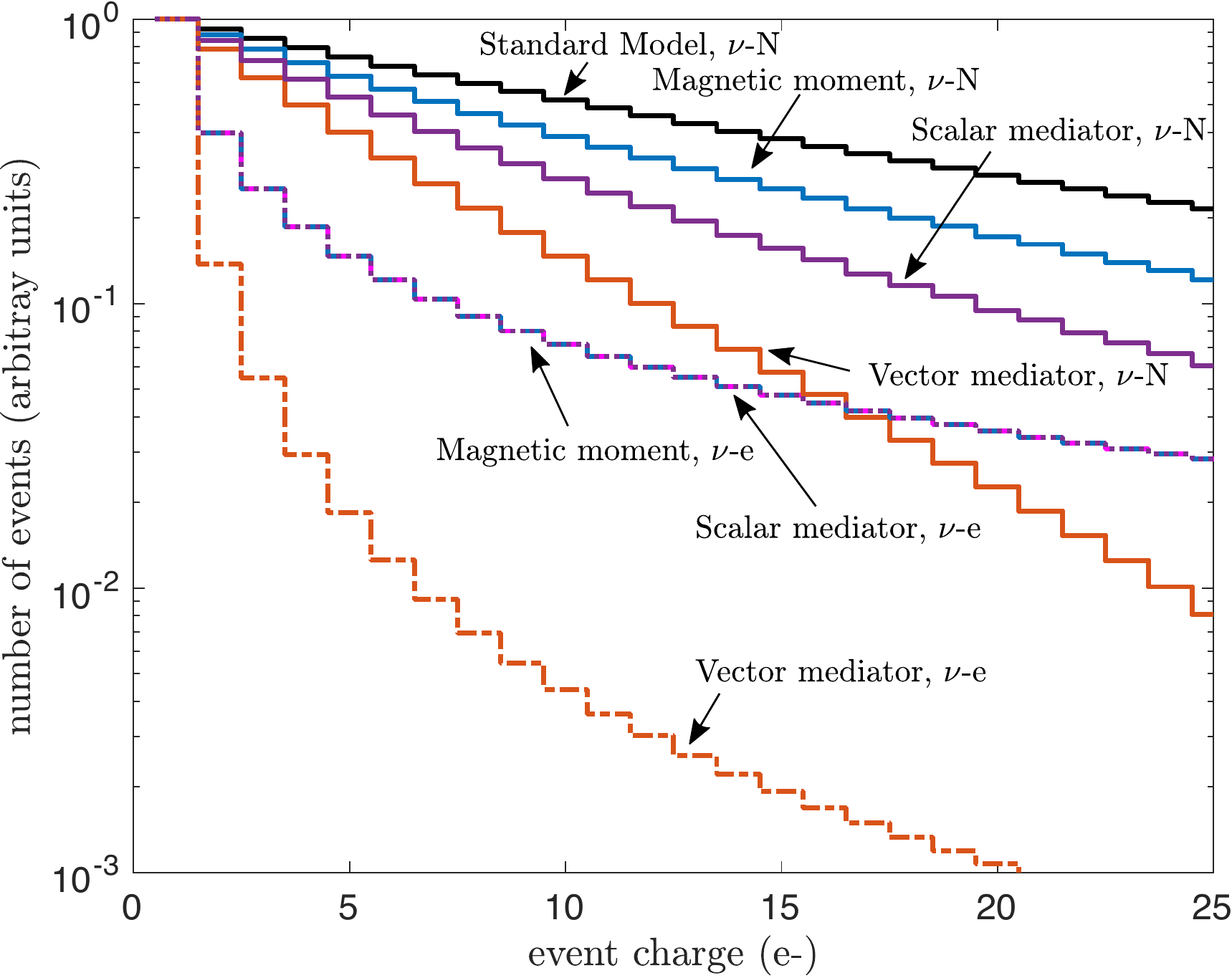} 

\caption{Charge distribution spectra for several different interactions and scattering channels normalized by the first bin height. The interaction parameter values used for magnetic moment $\nu$-$N$ and $\nu$-$e$: $\mu_{\nu_e}=8\times10^{-11}\mu_B$,
for vector mediator, $\nu$-$N$: $m_{Z^{\prime}} = 1\times10^{-6}$ MeV, $\sqrt{g_{\nu Z^{\prime}}g_{qv}} = 1.2\times10^{-5}$.
for vector mediator, $\nu$-$e$: $m_{Z^{\prime}} = 1\times10^{-6}$ MeV, $\sqrt{g_{\nu Z^{\prime}}g_{lv}} = 4\times10^{-7}$.
for scalar mediator, $\nu$-$N$: $m_{\phi} = 1\times10^{-6}$ MeV, $\sqrt{g_{\nu\phi}g_{qs}} = 2\times10^{-6}$.
for scalar mediator, $\nu$-$e$: $m_{\phi} = 1\times10^{-6}$ MeV, $\sqrt{g_{\nu\phi}g_{ls}} = 4\times10^{-6}$. }
    \label{fig:charge_spectrum}
\end{figure}

Note that we have picked mediator masses sufficiently low to make the impact of low thresholds more evident. This is particularly true for vector mediators, as the differential cross section goes as $\propto1/E_R^2$, if for light enough masses. In fact, more than 99\% of the events due to new physics takes place below 1 keV recoil energy for all our benchmarks. Besides the low recoil energy enhancement coming from the lightness of new physics, the quenching factor for neutrino-nucleus scattering reduces the ionization energy with respect to the true recoil of the nucleus, further evidencing the advantage of low energy thresholds. This is more obvious in Fig.~\ref{fig:charge_spectrum}, where we present the normalized neutrino-electron and neutrino-nucleon charge distribution spectra for several BSM scenarios. Clearly, $\nu-e$ scattering via light vector mediators is the scenario which benefits the most from low ionization thresholds while magnetic moment $\nu$-nucleus scatter is the most ``flat'' spectrum. Note that we show the spectra regardless of the detection threshold for presentation purposes. Quantum effects including  electron energy levels were not included~\cite{Essig:2015cda}.

\subsection{Forecasted Sensitivity to Neutrino Magnetic Moment and New light mediators}

Now we present the experimental sensitivity of our setup to light mediators and a non-standard neutrino magnetic moment. For simplicity, we will consider that the light mediator couples either to electrons or to quarks. We will assume an experimental benchmark configuration with the Chavarria quenching, a 5\% reactor neutrino flux normalization uncertainty, and 1~kdru flat background rate.
For neutrino-electron scattering, the dominant uncertainty that drives the sensitivity is the reactor flux. Due to that, we will also present, for this case, the results assuming the data-driven flux inferred by the Daya Bay measurements, as well as a statistics only curve for reference.
These systematics have little impact on the neutrino-nucleus results, so we will only show the main experimental benchmark in that case. For the coupling to electrons, we will compare our sensitivity to the XENON1T excluded region obtained in Ref.~\cite{Boehm:2020ltd}. For the case of coupling to quarks, we also show current limits from CONNIE~\cite{Aguilar-Arevalo:2019zme} and  COHERENT~\cite{Liao:2017uzy, Khan:2019cvi}. Limits on the pseudo-scalar and axial scenarios coupling with electrons coming from GEMMA~\cite{Beda:2013mta} and TEXONO~\cite{Deniz:2009mu} were derived following the procedure established in~\cite{Lindner:2018kjo}. Briefly, we consider the last data related to neutrino-electron scatterings, and perform a spectral fit in terms of electron recoil energies. To the best of our knowledge such limits were not available in the literature until now.

\begin{figure}[h!]
    \centering
    \includegraphics[width = 0.45\textwidth]{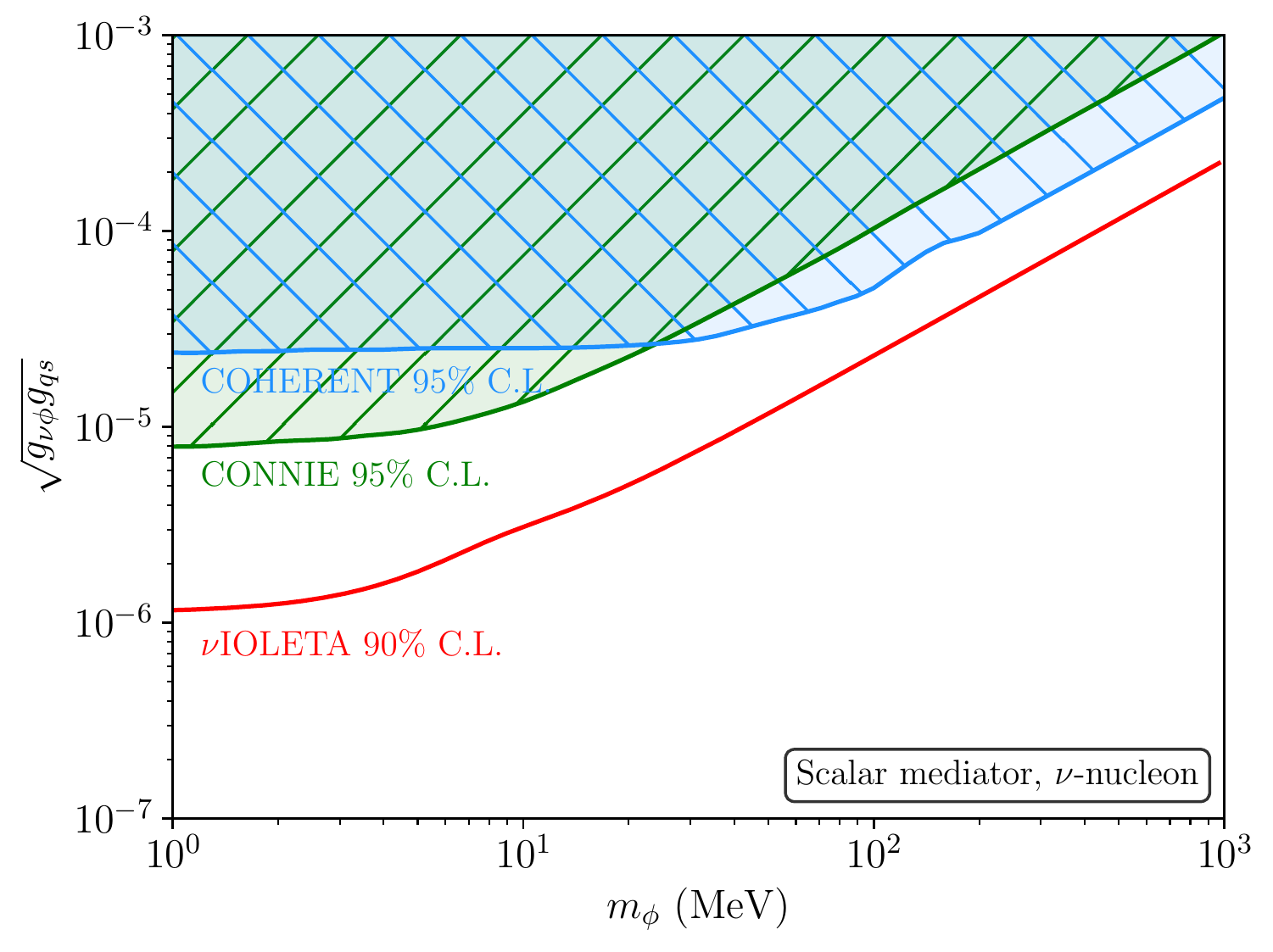}
    \includegraphics[width = 0.45\textwidth]{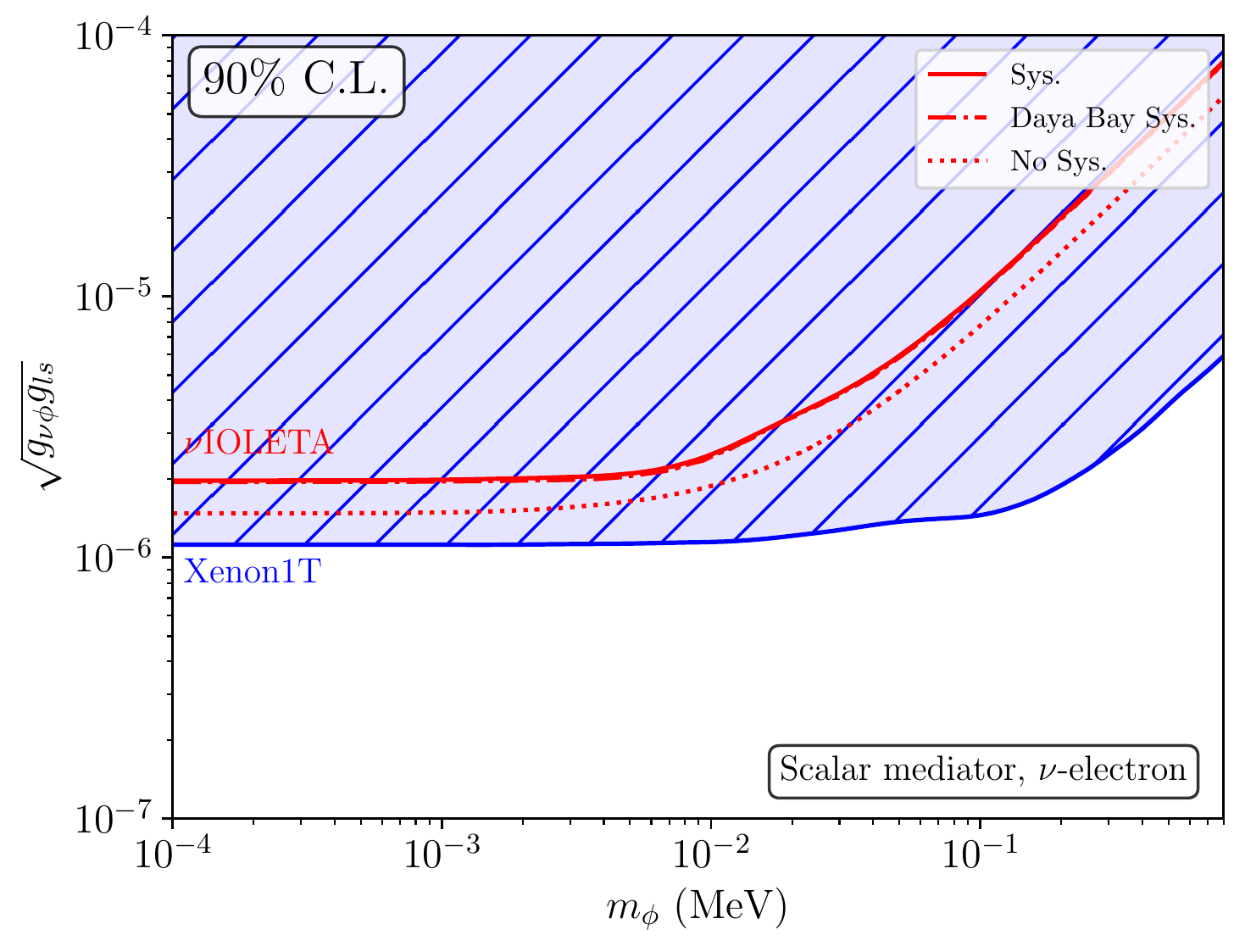}
    \caption{Sensitivity of the vIOLETA experiment (red) to a new scalar interaction between neutrinos and nucleons (left panel) and electrons (right panel). For the interaction with nucleons we compare with the bounds from the COHERENT (light blue) and CONNIE (green) experiments, while for electrons we compare with the recent bound from XENON1T (blue).
    For $\nu$-$e$ scattering, we also show the impact of systematic uncertainties: 5\% flux normalization (solid red), Daya Bay data-driven flux covariance matrix (dot-dashed), and only statistical uncertainties (dotted).}
    \label{fig:sens_lsm}
\end{figure}

In Fig.~\ref{fig:sens_lsm} we show the sensitivity at 90\%~C.L. to a new light scalar mediator coupling to neutrinos and quarks (left panel) or electrons (right panel). As can be observed in the left panel, vIOLETA (red) could improve current bounds by about an order of magnitude with respect to CONNIE (green) and a factor $\sim2.5$ with respect to COHERENT (light blue) for masses between 0.1-1~GeV. In the right panel we present the comparison between vIOLETA (red curves) and the  XENON1T (blue hatched region) exclusion of a light scalar mediator coupling to neutrinos and electrons. XENON1T is more competitive due to its larger detector mass and the fact that, for small $m_{\phi}$, the differential cross section goes as the inverse of the recoil energy. Therefore, the total cross section only increases logarithmically with a lower threshold: vIOLETA cannot take full advantage of its low-energy threshold capabilities in the light scalar mediator scenario.

\begin{figure}
    \centering
    \includegraphics[width = 0.45\textwidth]{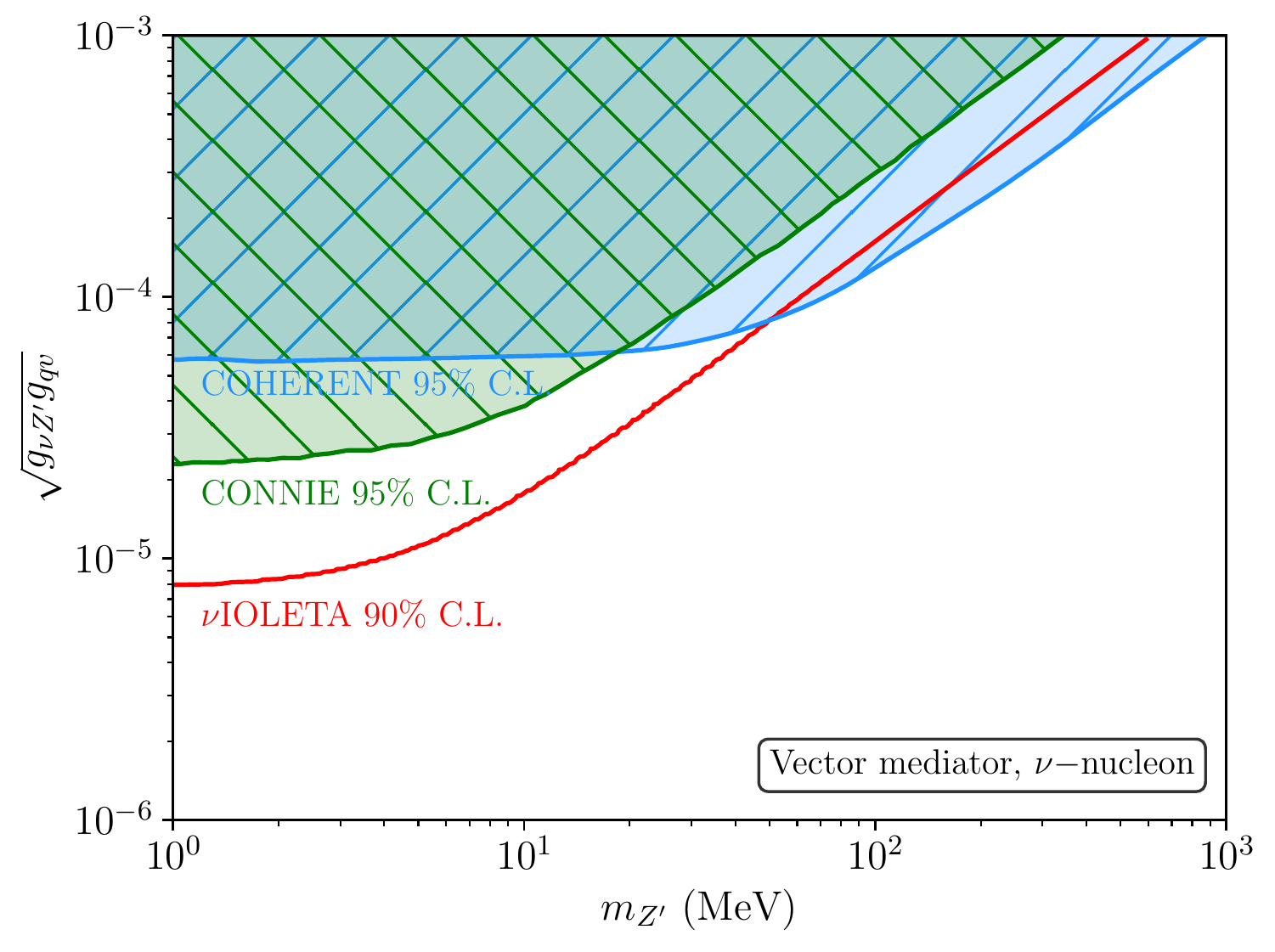}
    \includegraphics[width = 0.45\textwidth]{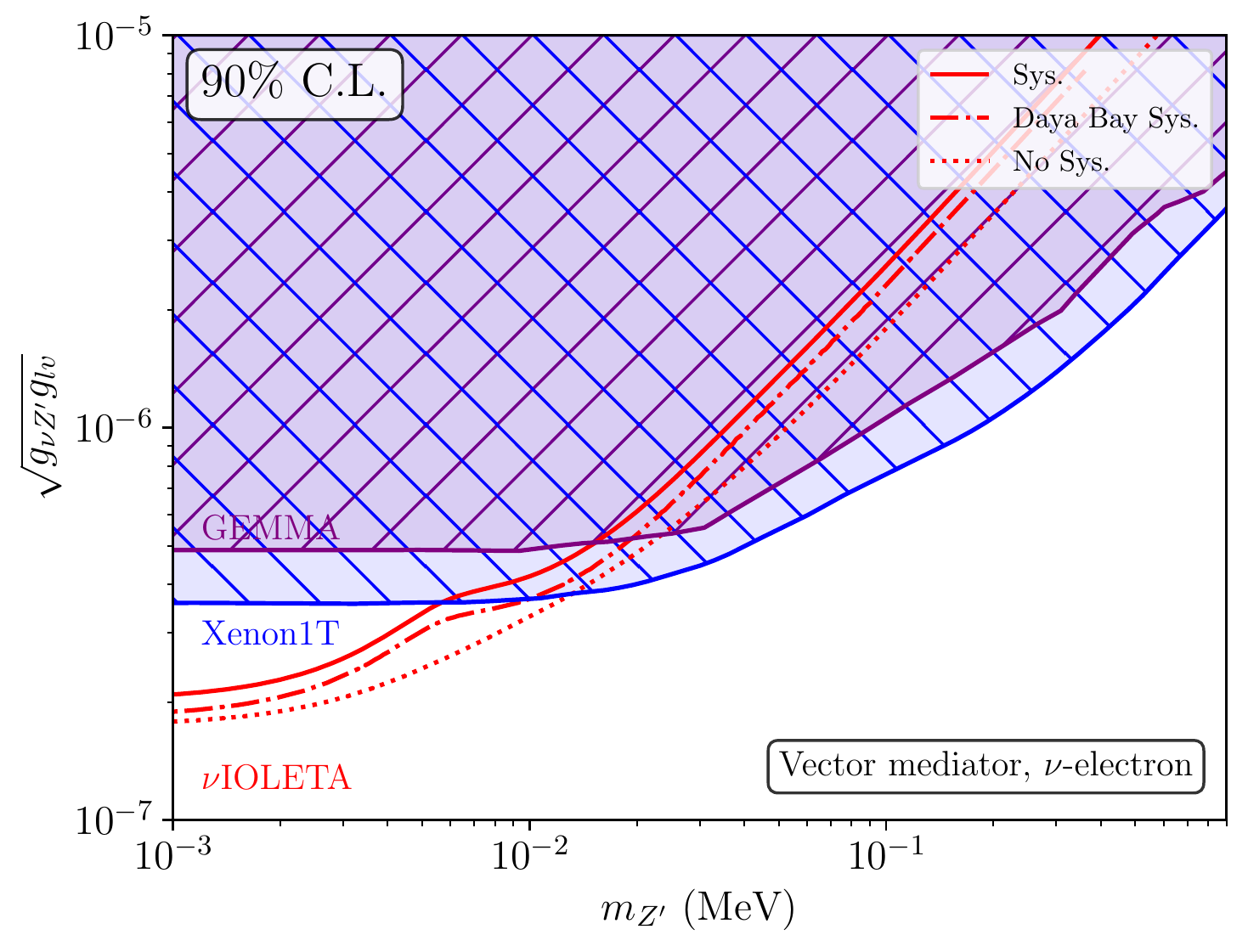}
    \caption{Sensitivity to a new interaction mediated by a light vector boson between neutrinos and nucleons (left panel) and electrons (right panel) as a function of the mass of the new boson and the coupling. We compare the sensitivity from vIOLETA (red) to the bounds from COHERENT (light blue) and CONNIE (green) for the interaction with nucleons and with Xenon1T (blue) and GEMMA (purple) for the case of an interaction with electrons. For $\nu$-$e$ scattering, we also show the impact of systematic uncertainties: 5\% flux normalization (solid red), Daya Bay data-driven flux covariance matrix (dot-dashed), and only statistical uncertainties (dotted). }
    \label{fig:sens_lvm}
\end{figure}

In Fig.~\ref{fig:sens_lvm} we show the sensitivity to a light vector mediator coupled to neutrinos and quarks (left panel) and electrons (right panel). Focusing on the left panel, we see that vIOLETA (red) can improve the current constraints for masses below about 50 MeV. Above this mass, COHERENT (blue) provides the best constraint. At low masses, vIOLETA may improve the leading bounds from CONNIE (green) by a factor 3 or so. For the coupling to electrons (right panel), vIOLETA (red) could improve the current sensitivity to masses below about 20 MeV by a factor $\sim2$. For larger masses, XENON1T would still have a better sensitivity due to the fact that at large $m_{Z^{\prime}}$ the cross section is independent of the recoil energy, as can be seen in Table~\ref{tab:Int}. Note also that the neutrino-electron scattering constraint exhibits some nontrivial feature around $M_{Z'}=6$~keV, which goes away when there are no flux systematics. The reason for this is because, at these masses, the neutrino-electron scattering induced by new physics leads to a ionization spectrum which is very similar to the standard (quenched) CEvNS spectrum. Therefore, any systematic uncertainty on the standard CEvNS spectrum causes a loss of sensitivity in that region. For lower masses, the $\nu-e$ spectrum is very peaked at low energies, leaving no room for confusion.

\begin{figure}
    \centering
    \includegraphics[width=0.45\textwidth]{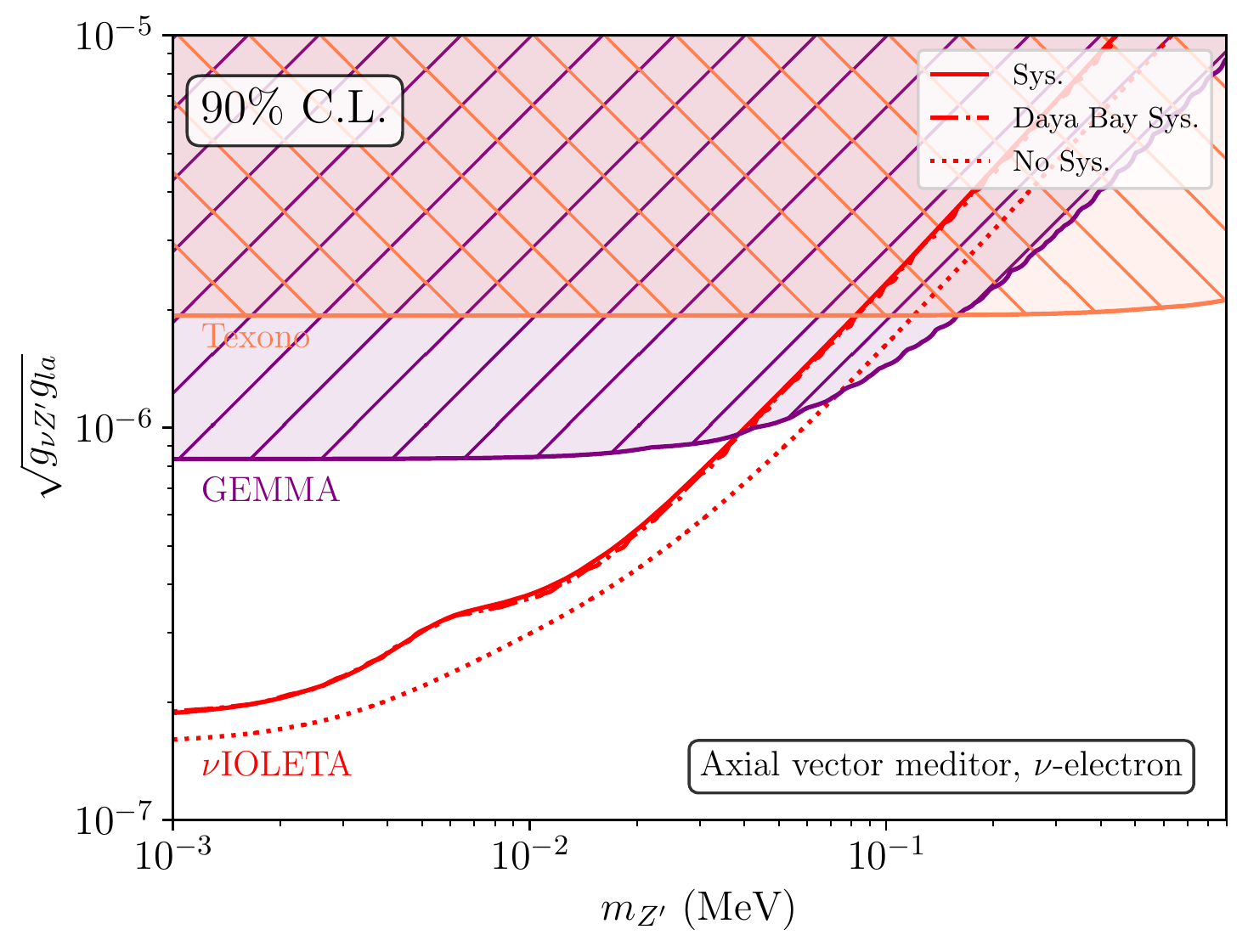}
    \includegraphics[width=0.45\textwidth]{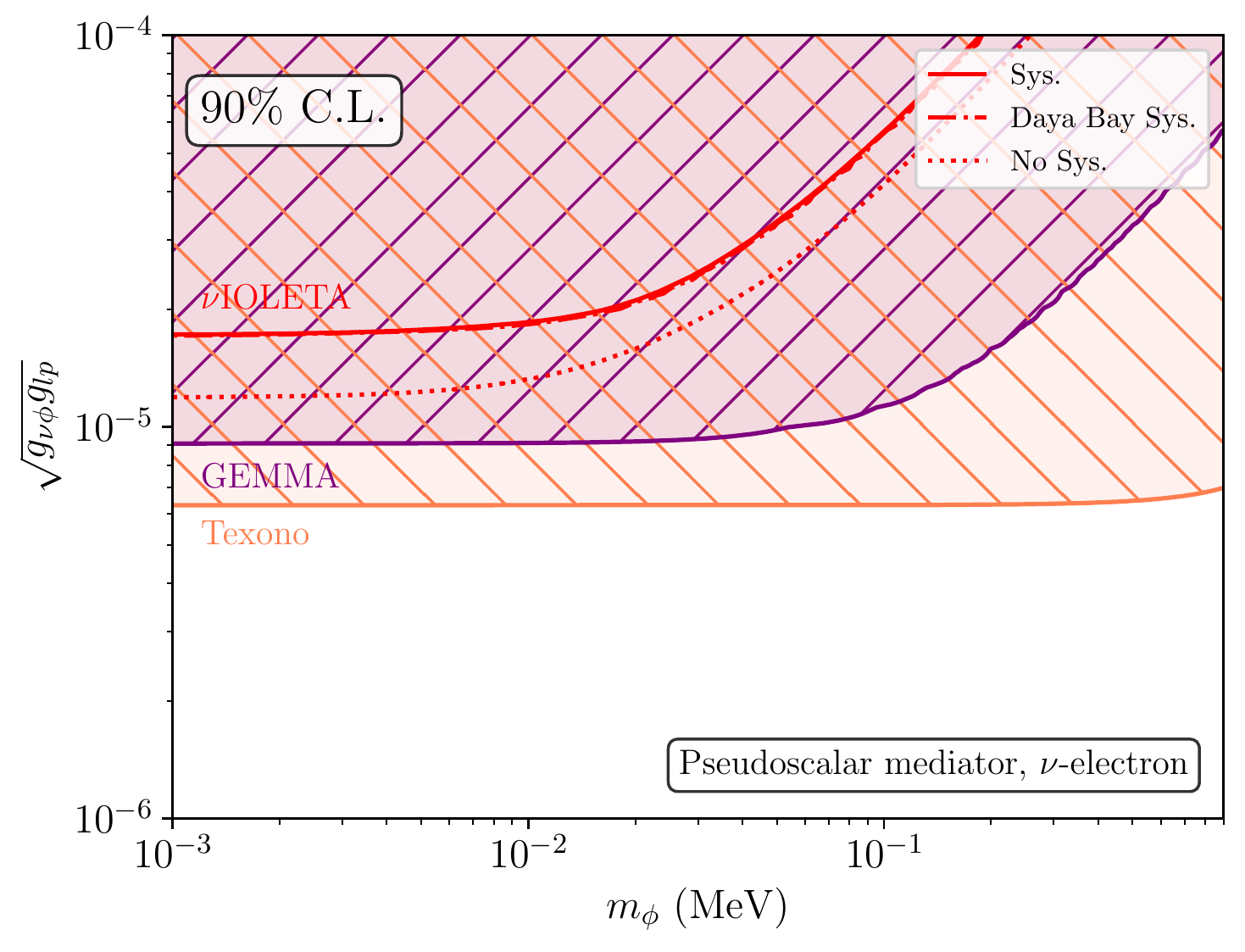}
    \caption{Sensitivity for the interaction with an axial vector (left) and a  pseudoscalar (right)   mediator coupling to neutrinos and electrons. We compare the vIOLETA sensitivity (red curves) to the bounds from GEMMA (purple) and Texono (coral). We also show the impact of systematic uncertainties: 5\% flux normalization (solid red), Daya Bay data-driven flux covariance matrix (dot-dashed), and only statistical uncertainties (dotted).}
    \label{fig:sens_avps}
\end{figure}

In Fig.~\ref{fig:sens_avps} we present the sensitivity for neutrino-electron scattering induced by axial (left panel)  and pseudoscalar (right panel) mediators. 
We do not show those for neutrino-nucleus scattering because pseudoscalars do not induce coherent neutrino-nucleus interactions, as they couple to the spin of the nucleus, while  limits on axial vector mediators coupling to neutrinos and nucleus are very similar to the vector mediator case.
We see that vIOLETA could improve the current GEMMA constraint on axial mediators by a factor $\sim4$ at low energies (left panel), while TEXONO dominates the pseudoscalar case for the region of interest.

For the neutrino magnetic moment, the future vIOLETA bound boils down to a single number, $\mu_{\nu_e}$. By performing the analysis we have found that 
\begin{equation}
  \mu_{\nu_e} < 5.3\times10^{-11}\mu_B\quad (90\%~{\rm C.L.})
\end{equation}
for the same experimental benchmark assumed in the light mediator scenarios.

\subsection{Dependence of the Forecasted Sensitivity on Experimental Assumptions}\label{sec:dependence-assumptions}

When deriving the future sensitivities to the light, weakly coupled new physics scenarios, we have made several assumptions on key experimental factors: recoil energy threshold, background rate, quenching factor, and systematic uncertainties. Although we have made an effort to assume a realistic experimental setup, it remains unclear what is the effect of such assumptions on the experimental sensitivity. As we have seen in the right panels of Figs.~\ref{fig:sens_lsm},~\ref{fig:sens_lvm} and~\ref{fig:sens_avps}, the uncertainty on the reactor neutrino flux is very relevant to the neutrino-electron scattering case. In fact, we have checked that the neutrino flux systematics is the dominant systematics in most cases, followed by the uncertainty on the fission fractions.

\begin{figure}[t]
    \centering
    \includegraphics[width = 0.45\textwidth]{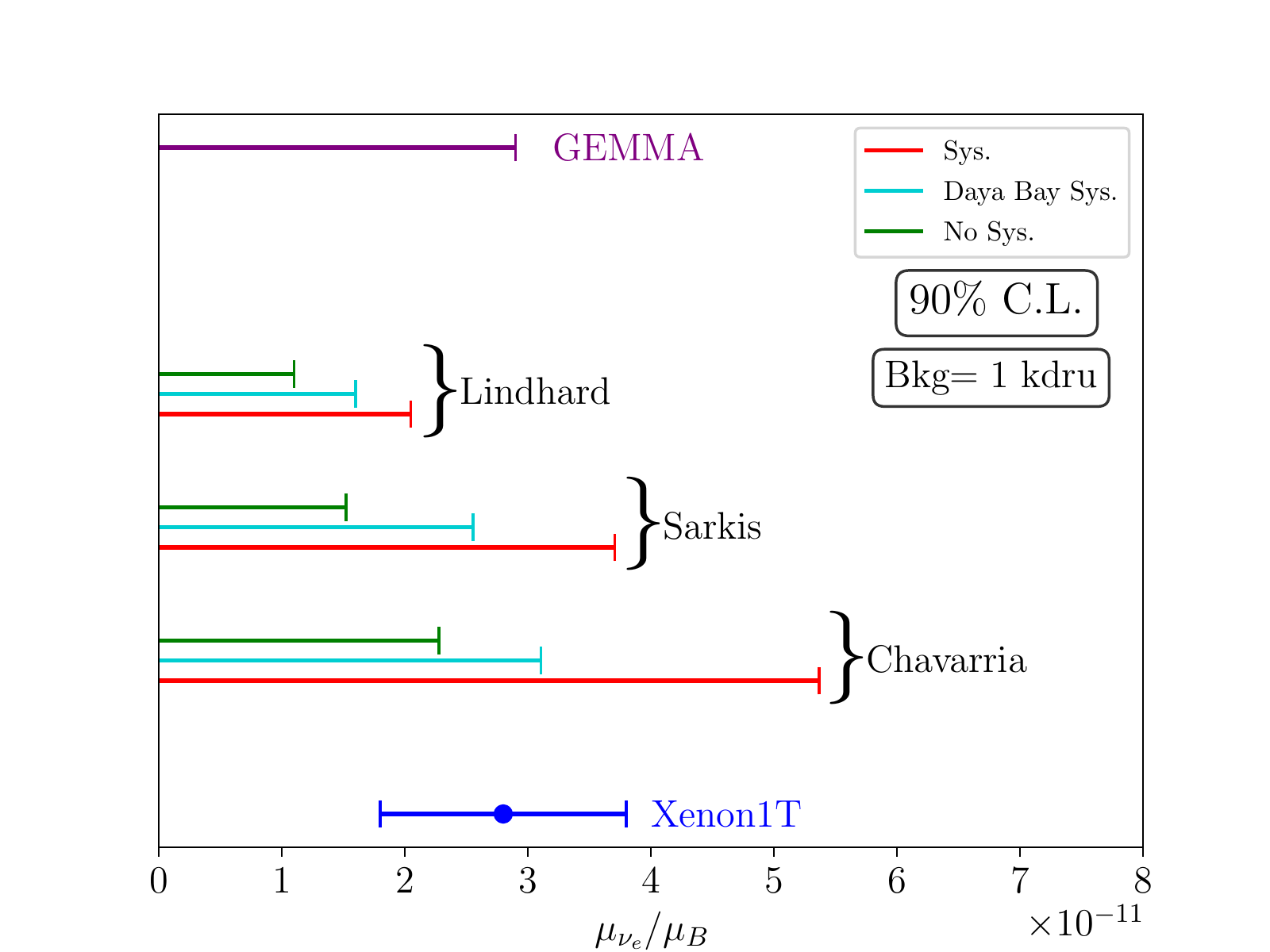}
    \includegraphics[width=0.45\textwidth]{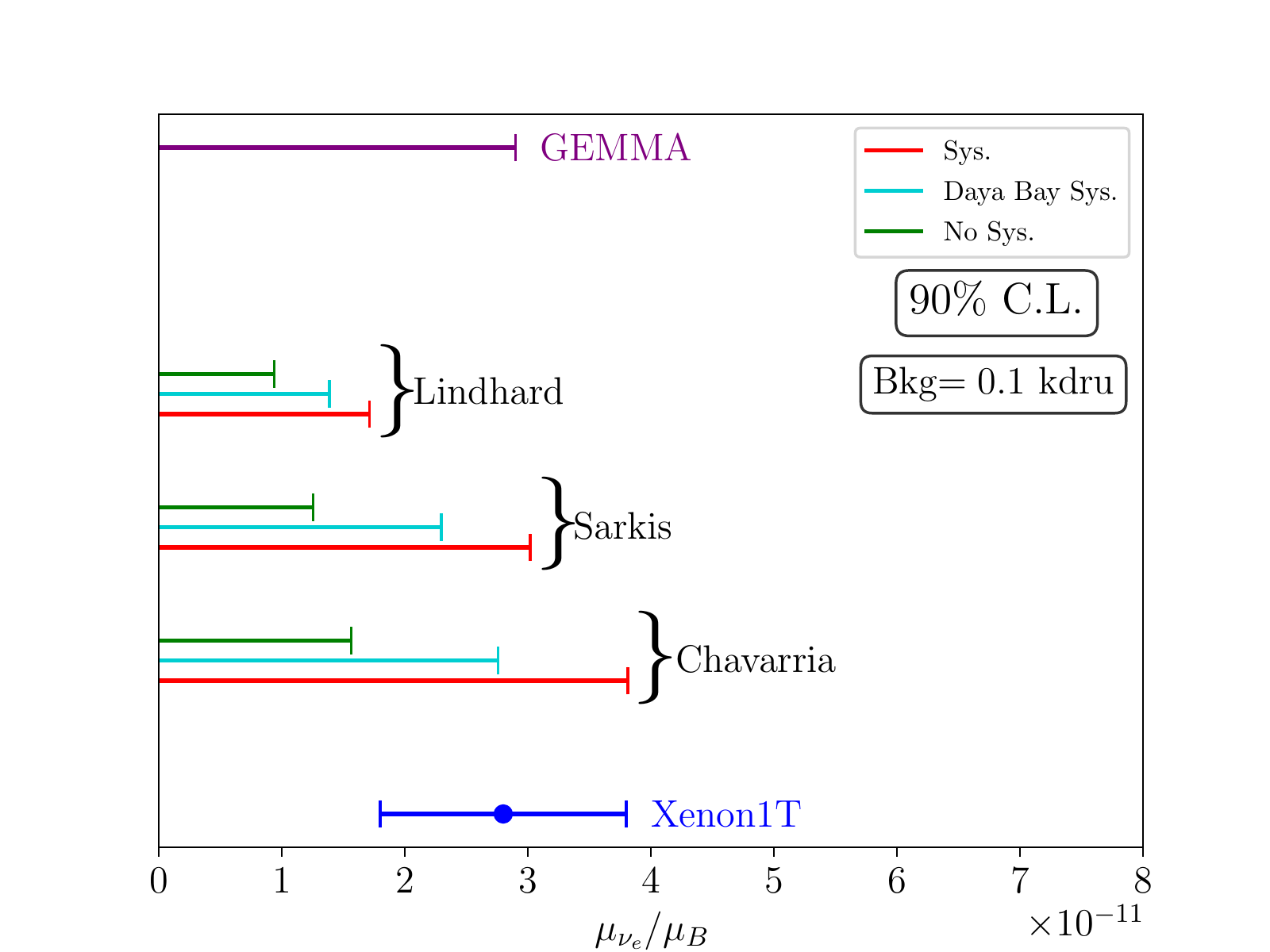}
    \caption{Sensitivity to non-standard neutrino magnetic moment at vIOLETA using Chavarria, Sarkis or Lindhard   quenching factors (as indicated in the figure), for a background of $1$~kdru (left panel) or $100$~dru (right panel). We also show the dependence of the sensitivity on the reactor neutrino flux uncertainty, where we show a 5\% overall systematics (red) and Daya Bay data-driven flux covariance matrix (cyan). We reference, we also show the sensitivity considering only statistical uncertainties (green). The purple line corresponds to the upper bound from GEMMA at $90\%$~C.L. and in blue we have the allowed region compatible with the Xenon1T result, at $90\%$~C.L.}
    \label{fig:sens_mm}
\end{figure}

To exemplify the impact of other assumptions in our results, we will present the neutrino magnetic moment sensitivity varying the quenching factor, the reactor flux uncertainty and the background rate. In Fig.~\ref{fig:sens_mm} we present such study. In the left panel we have assumed 10~kg detector mass and 1~kdru background rate all over, but we have varied the quenching factor assumption (Chavarria~\cite{Chavarria:2016xsi}, Sarkis~\cite{Sarkis:2020soy}  and Lindhard~\cite{lindhard}), as well as the flux systematic uncertainty between 5\% overall and the Daya Bay data-driven flux covariance matrix, following an analogous procedure to the one performed in~\cite{Fernandez-Moroni:2020yyl}. 
We also present, for each quenching factor assumption, the results obtained without any systematic uncertainties at all. This serves as a reference of the best ever attainable scenario given statistical uncertainties. The right panel is the same as the left one, except for a more optimistic background rate of 100~dru. This reduction on the background could be achieved with the Skipper-CMOS technology which improves on the timing resolution, allowing for active background veto.

As we can see, the role of the quenching factor is major for the sensitivity to a neutrino magnetic moment. The reason for that is simply statistics. If the quenching factor is larger, as in the case of Lindhard, more energy is deposited as ionization and thus a larger fraction of the reactor neutrino flux contribute to the  signal. We also see the relevance of measuring the reactor neutrino flux with inverse beta decay (IBD) experiments such as Daya Bay. Although the IBD threshold prohibits a measurement of the neutrino flux below the IBD threshold of 1.8~MeV, the data-driven knowledge of the flux above this threshold significantly boosts the sensitivity of vIOLETA. In the future, one would expect JUNO's near detector~\cite{Abusleme:2021zrw} to play an even more important role in this regard. The role of backgrounds is also quite clear from comparing the left and right panels, as we see a $\sim30\%$ improvement for almost all cases considered. We also see from these two figures that vIOLETA could significantly improve current constraints, and perhaps even probe the interpretation of the  XENON1T excess in terms of a light vector mediator or non-standard neutrino magnetic moment.

As a last remark, we would highlight the role of the low energy recoil threshold of vIOLETA in probing new physics. Skipper-CCDs are capable of counting single electrons with a readout noise as low as desired. If we, for instance, assume a one-sample readout noise of $\sigma\sim2$ electrons, as the one reported by the CONNIE experiment using conventional scientific grade CCDs~\cite{Aguilar-Arevalo:2019zme}, after measuring the charge in each pixel 256 times using the Skipper mode, a final sub-electron readout noise of $\sigma\sim2/\sqrt{256}=0.125$ electron could be achieved. Although this indicates that Skipper-CCDs could go down to a zero-electron energy threshold~\cite{Barak:2020fql}, the situation above ground is more challenging~\cite{Moroni:2021lyt}. The dominant source of pixel occupancy is the cosmic radiation (mainly atmospheric high energy particle showers), which is proportional to the exposure time. As a result, a typical exposure time of about a couple of hours corresponds to an expected pixel occupancy of the CCD of around 10-15~\cite{Aguilar-Arevalo:2019jlr}. In addition, photons produced by high energy cosmic particles via Cherenkov and bremsstrahlung processes would also contribute to the one-electron event rate as their energies could lead to ionized electrons in silicon~\cite{du2020sources}. Besides, there is a contribution due to the dark current, but is quite low in this kind of sensor ($\sim$10$^{-2}$ e$^-$/pix/day). All these background contributions grow linearly with exposure time. On the other side, the readout noise contributions are independent of time, being constant for each image. Therefore, the longer the exposure time the smaller the readout noise contribution.

Therefore, the silicon occupancy issue is essentially an optimization problem among cosmic backgrounds, readout noise and dark current. The nonzero occupancy caused by the effects discussed above would lead to an overestimate of the electron distribution at and near zero. Thus, even for pixel reads with such a good $\sigma$ the probability of misclassifying an empty pixel as one with charge is not zero. If we integrate this effect over hundreds of images (millons of pixels), a significant fraction of empty pixels will be classified as occupied. Two and three electron events could also appear in our images as results of spatial coincidences due to all aforementioned effects. However, the rate of fake multi-electron events is strongly suppressed. To be on the safe side, we have considered up to now that four electrons --which corresponds to 15~eV--, would be a conservative estimate of the experimental threshold for a Skipper-CCD experiment above surface. Nevertheless, a proper estimate and optimization of these effects could, in principle, allow the experiment to significantly lower its ionization threshold~\cite{Moroni:2021lyt}.

\begin{figure}[t]
    \centering
    \includegraphics[width = 0.45\textwidth]{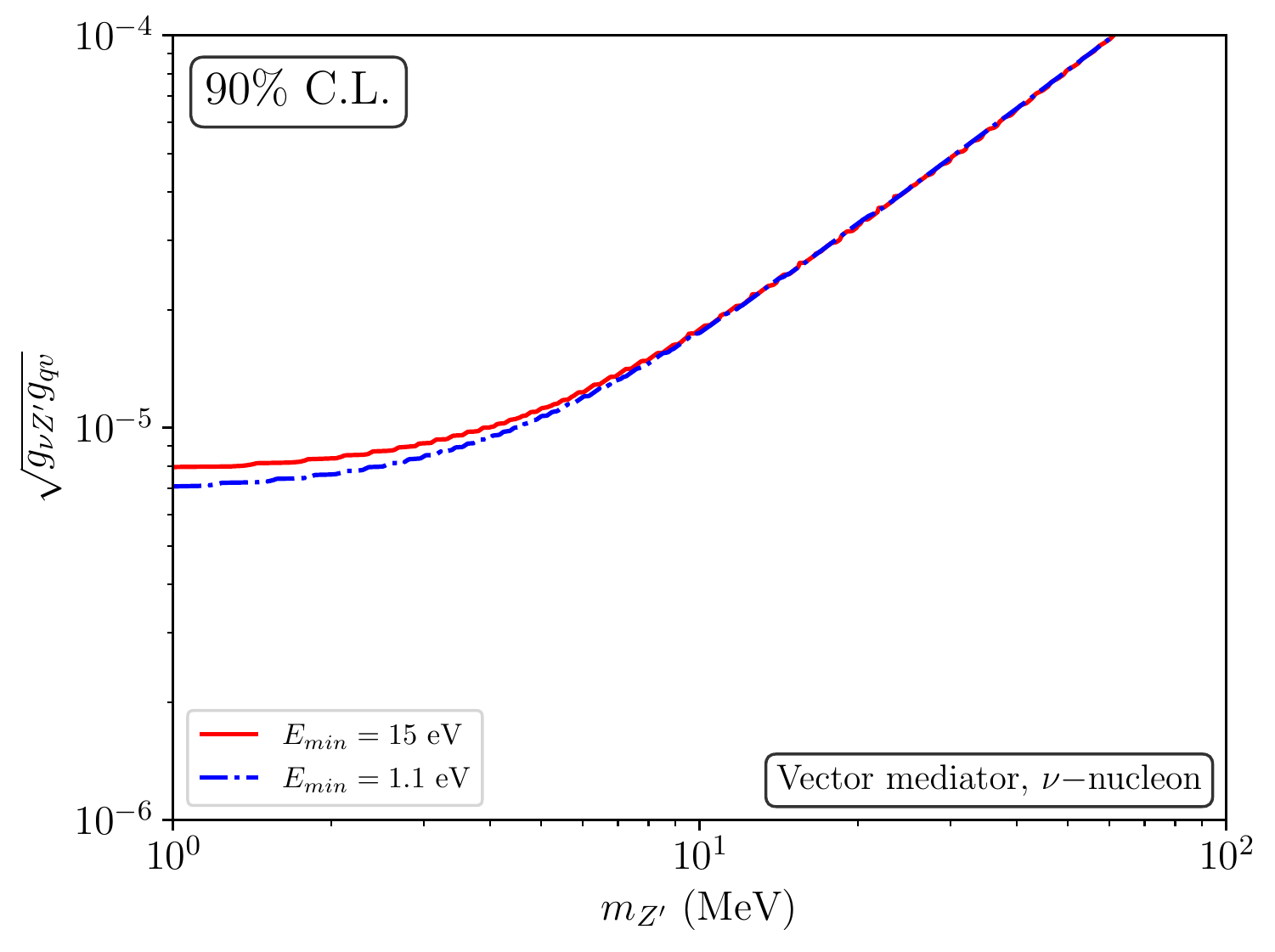}
    \includegraphics[width = 0.45\textwidth]{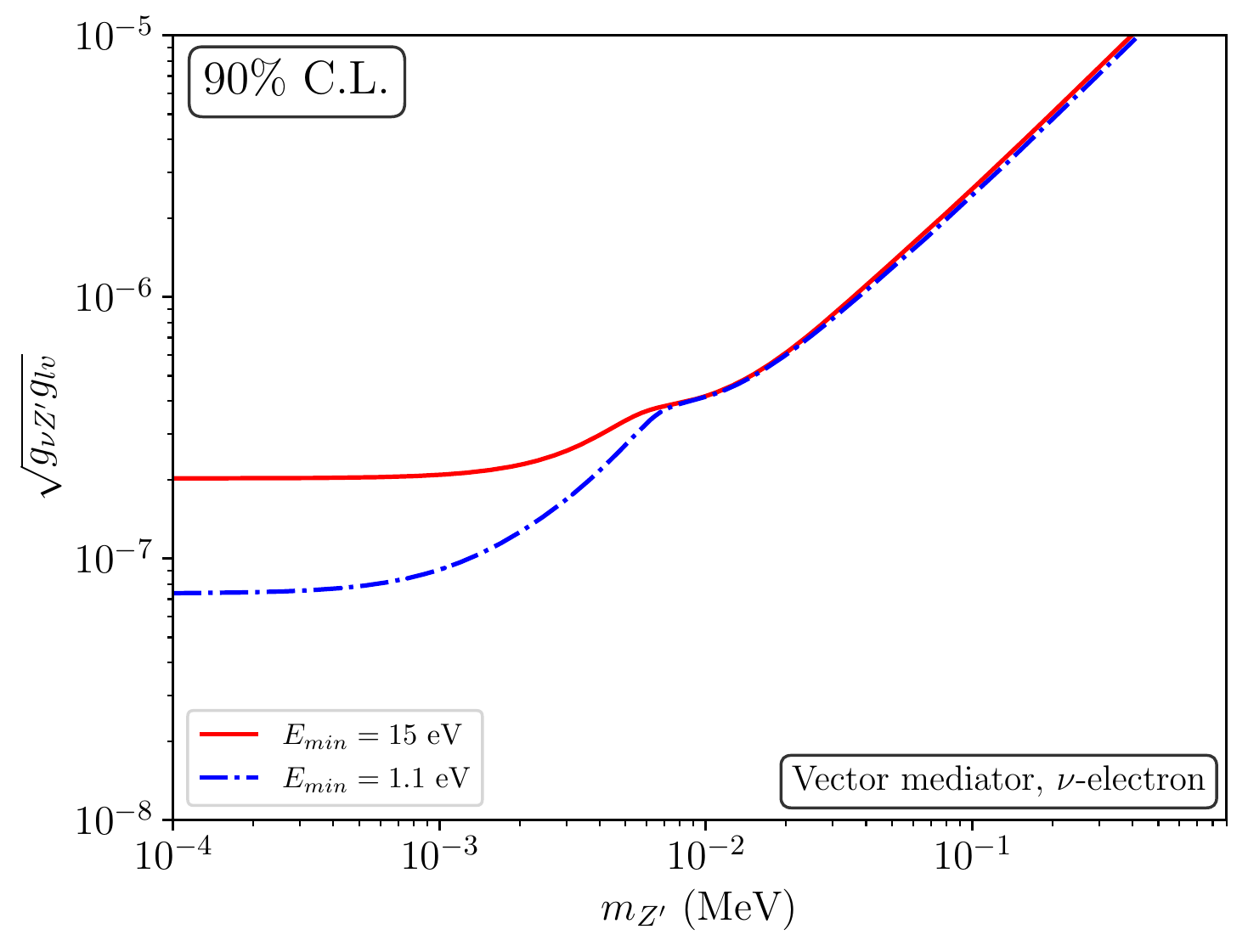}
    \caption{Comparison of vIOLETA's sensitivity to a new interaction between neutrinos and nucleons (left panel) or electrons (right panel) mediated by a light vector boson for two different energy thresholds: a conservative corresponding to $15$~eV  (red) and the limiting case for a silicon Skipper-CCD devise corresponding to the silicon band-gap of 1.1~eV (blue).}
    \label{fig:sens_comp_threshold}
\end{figure}

In Fig.~\ref{fig:sens_comp_threshold} we show vIOLETA's sensitivity to a new interaction between neutrinos and nucleons (left panel) or electrons (right panel) due to a light vector mediator for two different threshold energies, the conservative one of $15$~eV used up to here and the best case scenario, $1.1$~eV, corresponding to the silicon band-gap. All other experimental assumptions are unchanged from Table~\ref{tab:benchmark}. As we can see, for the neutrino-electron case, a lower threshold could improve the sensitivity by a factor of 2 for light enough mediators. This is due to the strong dependence on the recoil energy in the cross section when $m_{Z^{\prime}}\ll 2 E_R m_e$. Thus, the cross section, and consequently the number of events, is greatly enhanced when $E_R\rightarrow 0$. In contrast, for heavier mediator masses $M_{Z'}^2\gg 2\tilde{E}_R m$, where $ \tilde{E}_R$ are typical recoil energies and $m_N$ is the mass of the recoiled particle, the sensitivity does not depend on the threshold energy. This is expected, as the cross section becomes independent of the recoil energy. For the case of the interaction with nucleons there is not such an improvement for the different threshold energies. This can again be understood because now for small mediator masses the cross section is inversely proportional to $2 E_R m_N$ which does not give a large enhancement due to the large mass of silicon $m_N$.

\section{Conclusions}
\label{section:conclusions}

In this paper we have evaluated the sensitivity of the vIOLETA experiment --a 10~kg Skipper-CCD detector deployed 12~meters from a commercial nuclear reactor core-- to light, weakly coupled, beyond standard model scenarios. We have shown that, under reasonable assumptions, vIOLETA can improve current constraints on scalar and vector mediators coupling to neutrinos and quarks, as well as vector and axial vector mediators coupling to neutrinos and electrons. These improvements can be of up to almost an order of magnitude on the new physics couplings, particularly when the low threshold of Skipper-CCD detectors is leveraged (for very light mediators).

We have also provided a detailed study on the dependence of the experimental sensitivity on several assumptions regarding the experimental setup: systematic uncertainties, background rate, quenching factor, and ionization threshold. We have found that all these factors play a significant and comparable role in determining the experimental sensitivity to new physics. We highlight the important impact of unfolding the reactor neutrino flux from experimental measurements, such as the data-driven flux covariance matrix provided by the Daya Bay collaboration. Under optimistic but still realistic assumptions, vIOLETA can rule out the explanation of the  recent XENON1T excess in terms of a non-standard neutrino magnetic moment. We hope that this work will motivate the collaboration to search for light, weakly coupled new physics scenarios.


\acknowledgements
Fermilab is operated by the Fermi Research Alliance, LLC under contract No. DE-AC02-07CH11359 with the United States Department of Energy. This project has received support from the European Union’s Horizon 2020 research and innovation program under the Marie Skłodowska-Curie grant agreement No. 860881-HIDDeN. SR acknowledges support of the Spanish Agencia Estatal de Investigaci\'on and the EU ``Fondo Europeo de Desarrollo Regional'' (FEDER) through the projects PID2019-108892RB-I00/AEI/10.13039/501100011033 and of IFT Centro de Excelencia Severo Ochoa SEV-2016-0597.

\bibliography{CEvNS.bib}

\end{document}